\documentclass[a4paper,12pt,cite,epsfig]{article}
\pdfoutput=1
\usepackage{cite}
\usepackage[english]{babel}
\usepackage{hyperref}
\hypersetup{
  colorlinks   = true, %Colours links instead of ugly boxes
  urlcolor     = blue, %Colour for external hyperlinks
  linkcolor    = blue, %Colour of internal links
  citecolor   = red %Colour of citations
}
\usepackage{tocloft}

\usepackage{soul}
\usepackage{empheq}
\usepackage{ragged2e}%for alignment: see  https://www.sharelatex.com/learn/Text_alignment
\usepackage{url}
\usepackage[utf8]{inputenc}
\usepackage[toc,page]{appendix}
\usepackage[autostyle]{csquotes}
\usepackage{slashed}
\usepackage{amssymb}%for changing item shape into \barwedge, \boxdot, \boxminus, \boxplus, \boxtimes, \Cap, \Cup (and many more), the arrow \leadsto, and some other symbols such as \Box and \Diamond, \checkmark \blacksquare \bigstar \Diamond
\usepackage{graphicx}
\usepackage{dirtytalk}% quote and double quote
\usepackage{longtable}
\usepackage{array}%This Package is for table
\usepackage{bm} %This Package is for math bold
\usepackage{amsmath}
\usepackage{nccmath}
\usepackage{cancel}
\usepackage{mathtools}
\usepackage{amsfonts}
\usepackage{amssymb}
\usepackage{braket}
\usepackage[labelfont=bf]{caption}
\usepackage[a4paper,includeheadfoot,margin=2.54cm]{geometry}
\usepackage{hyperref}
\usepackage{cleveref}
\newcommand{\bse}{\begin{subequations}}
\newcommand{\ese}{\end{subequations}}
\numberwithin{equation}{section}
\begin{document}
\begin{titlepage}
\centering % Centre everything on the title page
	
	%\scshape % Use small caps for all text on the title page
	
	\vspace*{\baselineskip} % White space at the top of the page
	
	%------------------------------------------------
	%	Title
	%------------------------------------------------

	\vspace{0.5\baselineskip} % Whitespace above the title
	\flushleft{\LARGE{{\sffamily{\bfseries{Logarithmic correction to the entropy of extremal black holes in $\mathcal{N}=1$ Einstein-Maxwell supergravity}}}}} % Title
	
	%\vspace{0.75\baselineskip} % Whitespace below the title
	
	%\vspace{2\baselineskip} % Whitespace after the title block
\vspace{0.4in}	
	\rule{\textwidth}{1.6pt}% Thick horizontal rule
	%\rule{\textwidth}{0.4pt} % Thin horizontal rule
\vspace{0.4in}	
%\hspace{-0.5in}
\raggedright
{
\textbf{\small\scshape{Gourav Banerjee$^{1}$, Sudip Karan$^{2}$ and Binata Panda$^{3}$}}\\ \vspace{0.05in}
\textit{\footnotesize $^{1,2,3}$  Department of  Physics,\\ Indian Institute of Technology (Indian School of Mines),\\ Dhanbad, Jharkhand-826004, INDIA\\
Email: \href{mailto:gourav\_9124@ap.ism.ac.in}{$^{1}$gourav\_9124@ap.ism.ac.in}, \href{mailto:sudip.karan@ap.ism.ac.in}{$^{2}$sudip.karan@ap.ism.ac.in}, \href{mailto:binata@iitism.ac.in}{$^{3}$binata@iitism.ac.in}}}	
\vspace{0.7in}
\justify	
\begin{abstract}

We study one-loop covariant effective action of \say{non-minimally coupled} $\mathcal{N}=1$, $d=4$ Einstein-Maxwell supergravity theory by heat kernel tool. By fluctuating the fields around the classical background, we study the functional determinant of Laplacian differential operator following Seeley-DeWitt technique of  heat kernel expansion in proper time. We then compute  the Seeley-DeWitt coefficients obtained through the expansion. A particular Seeley-DeWitt coefficient is used for determining the logarithmic correction to Bekenstein-Hawking entropy of extremal black holes using quantum entropy function formalism. We thus determine the logarithmic correction to the entropy of Kerr-Newman, Kerr and Reissner-Nordstr\"{o}m black holes in {\say{non-minimally coupled}} $\mathcal{N}=1$, $d=4$ Einstein-Maxwell supergravity theory.
	
%\begin{center}

%\end{center}
\end{abstract}
%\rule{\textwidth}{0.4pt}\vspace*{-\baselineskip}\vspace{3.2pt} % Thin horizontal rule
	%\rule{\textwidth}{1pt} % Thick horizontal rule
\end{titlepage}
%\maketitle
%-------------------------------------------------------------------------------------
%                                                          ABSTRACT
%-------------------------------------------------------------------------------------

%\tableofcontents
\pagebreak 
%-----------------------------------------------------------------------------
 %\rule{\textwidth}{.5mm} 
	 \begingroup
	\hypersetup{linkcolor=black} 
	\tableofcontents
	\noindent\hrulefill
	\endgroup

\pagebreak 
%-----------------------------------------------------------------------------
 %\rule{\textwidth}{.5mm} 
   
\section{Introduction}\label{intro}
%\begin{ceqn}
Studying the spectrum of effective action is one of the exciting topics in quantum field theory at all times as it  contains the relevant information of any theory. It provides the knowledge of physical amplitudes and quantum effects of the fields present in the theory. The analysis of effective action by perturbative expansion of fields in various loops is a powerful way to study a particular theory. Heat kernel proves to be an efficient tool for analysis of effective action  at one-loop level \cite{Avramidi:1994th,Avramidi:2001ns}. The basic principle of this heat kernel method is to study an asymptotic expansion for the spectrum of the differential operator of effective action. It helps us in the investigation of one-loop divergence, quantum anomalies, vacuum polarization, Casimir effect, ultraviolet divergence, spectral geometry, etc. Within heat kernel expansion method, there are various approaches\cite{Avramidi:1994th,Avramidi:2001ns,Avramidi:1986mj,Mckean:1967,Avramidi,Avramidi2,VandeVen:1997pf,Fliegner:1994zc,Fliegner:1997rk,Gopakumar:2011qs,David:2009xg,Barvinsky:2017mal,Vassilevich:2003ll,Seeley:1967ea,DeWitt:1965ff,DeWitt:1967gg,DeWitt:1967hh,DeWitt:1967ii,Seeley:1966tt,Seeley:1969uu,Duff:1977vv,Christensen:1979ww,Christensen:1980xx,Duff:1980yy,Birrel:1982zz,Gilkey:1984xy,Peixoto:2001wx,Gilkey:1975cd} to analyze the effective action such as covariant perturbation expansion, higher derivative expansion, group theoretic approach, diagram technique, wave function expansion, Seeley-DeWitt expansion, etc.\footnote{For a general review regarding several approaches of heat kernel expansion, one can refer \cite{Vassilevich:2003ll}. } However, among them, Seeley-DeWitt expansion \cite{Vassilevich:2003ll} is better over other methods  for various reasons stated here. The other methods possess either some model dependency or background geometry dependency. Whereas the Seeley-DeWitt expansion is more general and standard. This neither depends upon background geometry nor on supersymmetry and is applicable for all background field configuration. These properties make the Seeley-DeWitt approach more appropriate on those manifolds where other approaches fail to work. In the regime of black hole physics, the heat kernel expansion coefficients are found to be very useful for computing the logarithmic corrections to the entropy of black holes \cite{Banerjee:2010qc,Banerjee:2011jp,Sen:2011ba,Bhattacharyya:2012ss,Karan:2019gyn,Charles:2015nn,Castro:2018hsc,Sen:2012rr,Keeler:2014bra,Larsen:2014bqa,Sen:2012dw}  and the current paper will mainly focus on this direction.
% We present an explicit review on the Seeley-DeWitt expansion approach  in section \ref {Heat kernel Technique}. In this work, we compute Seeley-DeWitt coefficients for {\say{non-minimally coupled}} $\mathcal{N}=1$ Einstein-Maxwell supergravity theory (EMSGT) in four dimensions following the work of Vassilevich 
%\cite{Vassilevich:2003ll}.

%Here, we implement the heat kernel tool for studying the one-loop effective action of locally supersymmetric $\mathcal{N}=1$ $\text{EMSGT}$ in four dimensions on a compact Riemannian manifold without boundary. We follow  Seeley-DeWitt expansion approach of the heat kernel for computing the Seeley-DeWitt coefficients ($a_{2n}$). Then we use a particular coefficient $a_4$ for determining  logarithmic correction to the entropy of extremal black holes in this theory. 

This paper is motivated by a list of reported works \cite{Karan:2017txu,Bhattacharyya:2012ss,Karan:2019gyn,Charles:2015nn,Castro:2018hsc,Banerjee:2010qc,Banerjee:2011jp,Sen:2012rr,Sen:2011ba,Keeler:2014bra,Larsen:2014bqa,Sen:2012dw}. In \cite{Bhattacharyya:2012ss}, Sen et al. determined the logarithmic corrections of extremal Kerr-Newman type black holes in non-supersymmetric 4D Einstein-Maxwell theory using the quantum entropy function formalism \cite{Sen:2008yk,Sen:2009vz,Sen:2008vm}. Here the essential coefficient, $a_4$, is calculated following the general Seeley-DeWitt approach  \cite{Vassilevich:2003ll}. In our previous work \cite{Karan:2019gyn}, we have also followed the same approach as \cite{Bhattacharyya:2012ss} and computed the logarithmic correction to the entropy of extremal black holes in minimal $\mathcal{N}=2$ Einstein-Maxwell supergravity theory ($\text{EMSGT}$) in 4D. The computation of the heat kernel coefficients  for extended supersymmetric Einstein-Maxwell theory for $\mathcal{N}\geq2$ is presented in  \cite{Charles:2015nn,Castro:2018hsc} by Larsen et al. However, the authors have used the strategy of field redefinition approach within the Seeley-DeWitt technique to compute the required coefficients and obtained the logarithmic corrected entropy of various types of non-extremal black holes. In contrast, we are interested in the computation of logarithmic correction to the entropy of extremal black holes by using the general approach  \cite{Vassilevich:2003ll}  for the computation of required coefficients. In \cite{Banerjee:2010qc,Banerjee:2011jp,Sen:2012rr,Sen:2011ba}, the heat kernel coefficients are computed via the wave-function expansion approach and then used for the determination of logarithmic corrections to the entropy of different extremal black holes in $\mathcal{N}=2$,  $\mathcal{N}=4$, and $\mathcal{N}=8$ $\text{EMSGTs}$. Again, in \cite{Keeler:2014bra,Larsen:2014bqa},  on-shell and off-shell approaches have been used to determine the required coefficients.  In these wave function expansion, on-shell and off-shell methods, one should have explicit knowledge of symmetry properties of the background geometry as well as eigenvalues and eigenfunctions of Laplacian operator acting over the background. Whereas the Seeley-DeWitt expansion approach  \cite{Vassilevich:2003ll}, followed in \cite{Karan:2017txu,Bhattacharyya:2012ss,Karan:2019gyn} as well as in the present work, is applicable for any arbitrary background geometry.

Logarithmic corrections are special quantum corrections to the Bekenstein-Hawking entropy of black holes that can be completely determined from the gravity side without being concerned about the UV completion of the theory \cite{Banerjee:2010qc,Banerjee:2011jp,Sen:2011ba,Bhattacharyya:2012ss,Karan:2019gyn,Charles:2015nn,Castro:2018hsc,Sen:2012rr,Keeler:2014bra,Larsen:2014bqa,Sen:2012dw}.  In the context of quantum gravity, logarithmic corrections can be treated as the `infrared window' into the microstates of the given black holes. Over the last few decades, supergravities are the most intensively used platform  to investigate logarithmic corrections within string theory. Supergravity theories can be realized as the low energy gravity end of compactified string theories (mainly on Calabi-Yau manifold), and this makes logarithmic correction results in supergravities a more significant IR testing grounds for string theories.\footnote{An excellent comparative review of logarithmic corrections and their microscopic consistency in string theory can be found in \cite{Mandal:2010cj,Sen:2014aja} } Following the developments discussed in the previous paragraph, the analysis of the effective action in $\mathcal{N} = 1, d = 4$ EMSGT is highly inquisitive for the determination of logarithmic correction to the entropy of extremal black holes. In the current paper, we prefer to do this study via the heat kernel tool following the Seeley-DeWitt expansion approach \cite{Vassilevich:2003ll} and use the results into the framework of the quantum entropy function formalism \cite{Sen:2008yk,Sen:2009vz,Sen:2008vm}.  The attractor mechanism for various black holes in $\mathcal{N}=1$ supergravity is presented in \cite{Andrianopoli:2007rm}. In \cite{Ferrara}, the authors have studied local supersymmetry of $\mathcal{N}=1$ EMSGT in four dimensions. Moreover, we aim to 
find the logarithmic correction  to the  black hole entropy of various extremal black hole solutions such as Reissner-Nordstr\"{o}m, Kerr and Kerr-Newman black holes in a general class of $\mathcal{N}=1$ EMSGT in four dimensions. Our results will provide a strong constraint to an ultraviolet completion of the theory if an independent computation is done to determine the same  entropy  corrections  from the microscopic description.

In the current paper, we particularly consider a \say{non-minimal} $\mathcal{N}=1$ EMSGT{\footnote{For details of this theory, refer to Section \ref{Einstein-Maxwell theory in N eq 1 supergravity}.}} in four dimensions on a compact Riemannian manifold without boundary. Here the gaugino field of the vector multiplet is non-minimally coupled to the gravitino field of the supergravity multiplet through the gauge field strength. We fluctuate the fields in the theory around an arbitrary background and  construct the quadratic fluctuated (one-loop) action. We determine the functional determinant of one-loop action and then compute the first three  Seeley-DeWitt coefficients for this theory. The coefficients are presented in \eqref{total a2n}. Then employing quantum entropy function formalism, the particular coefficient $a_{4}$ is used to determine the logarithmic correction to Bekenstein-Hawking entropy for Kerr-Newman, Kerr and Reissner-Nordstr\"{o}m black holes in their extremal limit. The results are expressed in \eqref{KNFinal} to \eqref{RNFinal}. 

Earlier, the logarithmic correction to the extremal Reissner-Nordstr\"{o}m black hole entropy in two different classes of $\mathcal{N}=1, d = 4$  EMSGT (truncated from $\mathcal{N}=2$ and minimal) are predicted in \cite{Ferrara:2011qf}. In the ``truncated'' $\mathcal{N}=1$ theory, all the field couplings are considered in such a way that the $\mathcal{N}=2$ multiplets can be kinematically decomposed into $\mathcal{N}=1$ multiplets. For this truncation, one needs to assume a list of approximations (see sec. 4.1 of \cite{Ferrara:2011qf}). In the ``minimal'' $\mathcal{N}=1$ theory, the coupling term between gravitino (of supergravity multiplet) and gaugino (of vector multiplet) field is customized so that there exist minimal couplings between the supergravity multiplet and vector multiplets.\footnote{In general, there exist a non-minimal coupling term $\bar{\psi}F\lambda$ between gravitino $\psi$ (of supergravity multiplet) and gaugino $\lambda$ (of vector multiplets) via the vector field (of vector multiplets) strength $F$ in an $\mathcal{N}=1$ theory \cite{Ferrara}. The \say{minimally coupled} $\mathcal{N}=1$ class assumed a constant vector strength matrix $F$ \cite{Ferrara:2011qf}.} But in the present paper, we will proceed with a third and more general class of $\mathcal{N}=1$ theory -- the ``non-minimal'' $\mathcal{N}=1$ EMSGT, where no such approximations are assumed for field couplings, and the general non-minimal coupling between the supergravity and vector multiplet is also turned on. All the three classes of $\mathcal{N}=1$ theory are distinct and unable to reproduce each other's results because of the different choice of field couplings. Both the ``truncated'' and ``minimal'' $\mathcal{N}=1$ theories are approximated, while the ``non-minimal'' $\mathcal{N}=1$ EMSGT is a general one. In work \cite{Ferrara:2011qf}, logarithmic corrections for only extremal Reissner-Nordstr\"{o}m are achieved using $\mathcal{N}=2$ data and a particular local supersymmetrization process. On the other hand, our plan is to calculate logarithmic corrections to the entropy of extremal Reissner-Nordstr\"{o}m, Kerr and Kerr-Newman black holes by directly analyzing the quadratic fluctuated action of the general ``non-minimal'' $\mathcal{N}=1$ EMSGT. As discussed, our results are unique, novel and impossible to predict from \cite{Ferrara:2011qf} and vice versa.

The present paper is arranged as follows. In Section \ref{Heat kernel Technique}, we have briefly revised the heat kernel technique for the analysis of effective action and introduced the methodology prescribed by Vassilevich et al. \cite{Vassilevich:2003ll} for the computation of required Seeley-DeWitt coefficients.
We described the four dimensional \say{non-minimal} $\mathcal{N}=1$ Einstein-Maxwell supergravity theory and  the classical equations of motion for the action in
Section \ref{Einstein-Maxwell theory in N eq 1 supergravity}.
In Section \ref{Heat Kernels in N eq 1 EMSGT}, we carried out the heat kernel treatment to obtain the first three Seeley-DeWitt coefficients of the \say{non-minimal} $\mathcal{N}=1$, $d=4$ EMSGT. The coefficients for bosonic fields are summarized from our earlier work \cite{Karan:2019gyn}, whereas the first three Seeley-DeWitt coefficients for fermionic fields are explicitly computed.
In Section \ref{Application of Seeley-DeWitt coefficients in Logarithmic correction of extremal Black holes}, we first discussed a general approach to determine the logarithmic correction to entropy of extremal black holes using Seeley-DeWitt coefficients via quantum entropy function formalism. Using our results for the Seeley-DeWitt coefficient, we then calculate logarithmic corrections to the entropy Kerr-Newman, Kerr, Reissner-Nordstr\"{o}m extremal black holes in \say{non-minimal} $\mathcal{N}=1$, $d=4$ EMSGT. We summarize our results in 
Section \ref{summary}. Some details of our calculation are given in  Appendix \ref{Appendix: E and omega}.

%.......................................................................................................................
\section{Effective action analysis via heat kernel}\label{Heat kernel Technique}
In this section, we are going to discuss the one-loop effective action in 4D Euclidean spacetime using the Seeley-DeWitt expansion approach of heat kernel and review the basic calculation framework of this approach from \cite{Vassilevich:2003ll}. As initiated by Fock, Schwinger and DeWitt \cite{Fock:1937dy,Schwinger:1951nm,DeWitt:1965ff,DeWitt:1975ps}, a Green function is introduced and studied by the heat kernel technique for this analysis.
% in the analysis of effective action in four dimensional Euclidean spacetime .
The integral of this Green function over the proper time coordinate is a convenient way to study the one-loop effective action. The proper time method  defines the Green function in the neighborhood of light cone, which made it suitable for studying UV divergence and spectral geometry of effective action in one-loop approximation \cite{Avramidi:2000bm}.

%...............................................
\subsection{Loop expansion and effective action}

We begin with a  review of the heat kernel approach demonstrated in \cite{Vassilevich:2003ll} for studying the one-loop effective action by computing the functional determinant of quantum fields fluctuated around the background. We also analyze the relationship between effective action and heat kernels.\footnote{For an explicit elaboration, one can refer to \cite{Karan:2019gyn}.} 

We consider  a set of arbitrary fields $\varphi_{m}$\footnote{Note that this set also includes metric field.} on four dimensional compact, smooth Riemannian manifold without boundary. Then, 
the generating functional for Green function  with these   fields $\varphi_{m}$ in a Euclidean path integral representation\footnote{ The Euclidean path integral is followed by Wick rotation. Note that we have set $\hbar=c=1$ and  $G_N=\frac{1}{16 \pi}$ throughout the paper.} is expressed as
\begin{equation}\label{partition function}
	\mathcal{Z}= \int [D\varphi_{m}] \exp(-\mathcal{S}[\varphi_{m}]),
\end{equation}
\begin{equation}\label{S}
	\mathcal{S}[\varphi_{m}]=\int d^4 x \sqrt{g}\mathcal{L}(\varphi_{m},\partial_{\mu} \varphi_{m} ),
\end{equation}
where $[D\varphi_{m}]$ denotes the functional integration over all the possible fields $\varphi_{m}$ present in the theory. $\mathcal{S}$ is the Euclidean action with lagrangian density $ \mathcal{L}$ carrying  all information of fields in the manifold defined by metric $g_{\mu\nu}$.
We fluctuate the fields around any background solution to study the effect of perturbation in the action,  
\begin{equation}\label{F}
	\varphi_{m}=\bar{\varphi}_{m}+\tilde{\varphi}_{m},
\end{equation}
where $\bar{\varphi}_{m}$ and $\tilde{\varphi}_{m}$ symbolize the background and fluctuated fields, respectively. We are actually interested in the quantum characteristics of the action around a  classical background, which serves as a semi-classical system in an energy scale lower than the plank scale. For that, we consider the classical solutions $ \bar{\varphi}_{m}$ of the theory as a background and the Taylor expansion of the action (\ref{S})  around the classical background fields up to quadratic fluctuation fields yields,
\begin{equation}\label{A E}
	\mathcal{S} \approx \mathcal{S}_{\text{cl}}+\mathcal{S}_{2}=\mathcal{S}_{\text{cl}}+\int d^4 x \sqrt{\bar{g}} \tilde{\varphi}_{m}\Lambda^{mn}{\tilde{\varphi}}_{n}, 
\end{equation}
%\begin{equation}
%fkjnf
%\end{equation} where Scl is the classical action that only depends on the classical background. S2 denotes the quadratic interaction part of the quantum fields X in action via the differential operator \Lambda.
where $\mathcal{S}_{\text{cl}}$ is the classical action that only depends on the background. $\mathcal{S}_2$  denotes the quadratic interaction part of the quantum fields in action via the differential operator $\Lambda$.
%and depends upon geometrical invariant parameters.
% It is to note that the background fields and the fluctuated quantum fields are entirely different in nature.
% The Gaussian integral of partition function \ref{Z} around the string fields yields.
%\begin{equation}\label{Z f}
%\mathcal{Z}=exp(-\mathcal{S}_{cl}) det^{-1/2}(D)
%\end{equation}
After imposing the quadratic fluctuation (\ref{A E}), the Gaussian integral of generating functional (\ref{partition function}) gives the form of one-loop effective action $W$  as \cite{Avramidi:1994th,Peixoto:2001wx}
%Moreover, the Gaussian integral of generating functional (\ref{partition function}) by imposing the condition (\ref{A E}) over all  fields gives the one-loop effective action part $W$ for quadratic fluctuated quantum fields around the background, \cite{Avramidi:1994th,Peixoto:2001wx} %\footnote{
%We will follow the sign for fermions for the formulation of effective action is considered for the case of fermions only as we will mainly deal with fermionic multiplet in calculational part in this work, the bosonic contribution will be retrieved form our earlier works.}
\begin{equation}\label{W}
	W= \frac{\chi}{2}\ln \det (\Lambda)=\frac{\chi}{2} \text{tr} \ln \Lambda
	\enspace\forall 
	\begin{cases}
		\chi=+1 : \text{Bosons}\\
		\chi=-1 : \text{Fermions}
	\end{cases},
\end{equation}
where the sign of $\chi$ depicts the position of det$\Lambda$ in the denominator  or numerator for bosonic or fermionic field taken under consideration.\footnote{$\exp(-W)=\int{[D \tilde{\varphi}_m]\exp(-\int d^4 x \sqrt{\bar{g}}\tilde{{\varphi}}\Lambda \tilde {\varphi})}=\det^{-\chi/2}\Lambda$\cite{Castro:2018hsc}.} $W$ contains the detailed information of the fluctuated quantum fields  at one-loop level. In order to analyze the spectrum of the differential operator $\Lambda$ acted on the perturbed fields, we use the heat kernel technique by defining the  heat kernel $K$ as\cite{Avramidi:1994th,Vassilevich:2003ll}
\begin{equation}
	K(x,y;s)=\Bra{x}\exp(-s \Lambda)\Ket{y},
\end{equation}
where $K(x,y;s)$ is the Green function between the points $x$ and $y$ that  satisfies the heat equation,
\begin{equation}
	(\partial_{s}+\Lambda_{x}) K(x,y;s)=0,
\end{equation}
with the boundary condition, 
\begin{equation}
	K(x,y;0)=\delta (x,y).
\end{equation}
Here the proper time $s$ is known as the  heat kernel parameter. The one-loop effective action $W$ is related to the  trace of heat kernel, i.e., heat trace $D(s)$ via \cite{Vassilevich:2003ll,Karan:2019gyn}
\begin{equation}\label{W l}
	W=-\frac{1}{2} \int_{\epsilon}^{\infty} \frac{1}{s} ds \chi D(s),\enspace D(s)=\text{tr}~ (e^{-s\Lambda})= \sum_{i}e^{-s \lambda_{i}},
\end{equation}
where $\epsilon$ is the cut-off limit\footnote{ $\epsilon$ limit is controlled by the Planck parameter, $\epsilon\sim l_p^2\sim G_{N}\sim \frac{1}{16\pi}$.}  of effective action \eqref{W l} to counter the divergence in the lower limit regime and $\lambda_{i}$'s are non zero eigenvalues of $\Lambda$ which are discrete, but they may also be continuous.
$D(s)$  is associated with a perturbative expansion in proper time $s$ as  \cite{Vassilevich:2003ll,DeWitt:1965ff,DeWitt:1967gg,DeWitt:1967hh,DeWitt:1967ii,Seeley:1966tt,Seeley:1969uu,Duff:1977vv,Christensen:1979ww,Christensen:1980xx,Duff:1980yy,Birrel:1982zz,Gilkey:1984xy,Gilkey:1975cd}
\begin{equation}\label{G}
	D(s)=\int d^4 x \sqrt{\bar{g}}  K (x,x;s)=\int d^4 x \sqrt{\bar{g}} \sum_{n=0}^{\infty} s^{n-2} a_{2n}(x) .
\end{equation}
The proper time expansion of $K(x,x;s)$ gives a series expansion where the  coefficients $a_{2n}$,  termed as  Seeley-DeWitt coefficients \cite{Vassilevich:2003ll,DeWitt:1965ff,DeWitt:1967gg,DeWitt:1967hh,DeWitt:1967ii,Seeley:1966tt,Seeley:1969uu}, are functions of local background invariants such as Ricci tensor, Riemann tensor, gauge field strengths as well as covariant derivative of these parameters.
Our first task is to compute these coefficients $a_{2n}$\footnote{$a_{2n+1}=0$ for a manifold without boundary.} for an arbitrary background on a manifold without boundary. We then emphasize the applicability of a particular Seeley-DeWitt coefficient for determining the logarithmic correction to the entropy of extremal black holes.
%\begin{equation}
%K(t,D)=tr(e^{-sD})=\int d
%^4 x \sqrt{det {\bar{g}_{\mu\nu}}}
%\end{equation}
%.........................................................................................................................................
\subsection{Heat kernel calculational framework and methodology} 
\label{Heat kernel calculational framework and methodology}
%...................................................................................................
In this section,  we try to highlight the  generalized  approach for the computation of Seeley-DeWitt coefficients. For the four dimensional compact, smooth Riemannian manifold without boundary\footnote{The same methodology is also applicable for manifold with a boundary defined by some boundary conditions (see section 5 of \cite{Vassilevich:2003ll}).} under consideration, we will follow the procedure prescribed by Vassilevich  in  \cite{Vassilevich:2003ll} to compute the Seeley-DeWitt coefficients. We aim to express the required coefficients in terms of local invariants.\footnote{ Kindly refer our earlier works \cite{Karan:2017txu,Karan:2019gyn} for details on this discussion.}

We start by considering an ansatz $\bar{\varphi}_{m}$, a set of fields satisfying the classical equations of motion of a particular theory. We then fluctuate all the fields around the background as shown in \eqref{F} and the quadratic fluctuated action $\mathcal{S}_{2}$ is defined as
\begin{equation}
	\mathcal{S}_{2}=\int d^4 x \sqrt {\bar{g}}\tilde{\varphi}_{m} \Lambda^{mn}\tilde{\varphi}_{n}.
\end{equation}
We study the spectrum of elliptic Laplacian type partial differential operator $\Lambda^{mn}$ on the fluctuated quantum fields on the manifold. Again, $\Lambda^{mn}$  should be a Hermitian operator for the heat kernel methodology to be applicable. Therefore, we have, %\footnote{The sign is taken as per fermion methodology as discussed in \eqref{W} as our computation mainly deals with fermionic sector part and bosonic part is only taken from \cite{Bhattacharyya:2012ss}\textcolor{blue}{cite our paper also}},
\begin{equation}\label{E}
	\tilde{\varphi}_{m}\Lambda^{mn}\tilde{\varphi}_{n}=\pm\left[ \tilde{\varphi}_{m}\left(D^{\mu}D_{\mu}\right)G^{mn}\tilde{\varphi}_{n}+\tilde{\varphi}_{m}\left(N^{\rho} D_{\rho} \right)^{mn}\tilde{\varphi}_{n}+\tilde{\varphi}_{m}P^{mn}\tilde{\varphi}_{n} \right],
\end{equation}
where an overall +ve or -ve sign needs to be extracted out in the quadratic fluctuated form of bosons and fermions, respectively. $G$ is effective metric in field space, i.e., for vector field $G^{\mu\nu}=\bar{g}^{\mu\nu}$, for Rarita-Schwinger field $G^{\mu\nu}=\bar{g}^{\mu\nu}\mathbb{I}_{4}$ and for gaugino field $G=\mathbb{I}_{4}$. Here $\mathbb{I}_{4}$ is the identity matrix associated with spinor indices of fermionic fields.  $P$ and $N$ are arbitrary matrices, $D_{\mu}$ is an ordinary covariant derivative acting on the quantum fields. %All of them are constructed over the quantum fields.
Moreover, while dealing with a fermionic field (especially Dirac type) whose operator, say $O$ is linear, one has to  first make it  Laplacian of the form \eqref{E} %and then evaluate the functional determinant of the operator
using the following relation \cite{Peixoto:2001wx,Charles:2015nn,Sen:2011ba}
\begin{equation}\label{L}
	\ln \det (\Lambda^{mn})= \frac{1}{2} \ln \det({O^{mp}{{O_p}^n}^\dagger}).
\end{equation}
Equation \eqref{E} can be rewritten in the form below
\begin{equation}\label{V}
	\tilde{\varphi}_{m}\Lambda^{mn}\tilde{\varphi}_{n}=\pm\left(\tilde{\varphi}_{m}\left(\mathcal{D}_{\mu}\mathcal{D}^{\mu} \right)\mathbb{I}^{mn}\tilde{\varphi}_{n}+\tilde{\varphi}_{m}E^{mn}\tilde{\varphi}_{n}\right),
\end{equation}
where $E$ is endomorphism on the field bundle and $\mathcal{D}_{\mu}$ is a new effective covariant derivative defined with field connection $\omega_{\mu}$,
\begin{equation}\label{effective D}
	\mathcal{D}_{\mu}=D_{\mu}+\omega_{\mu}.
\end{equation}
The new form of \eqref{V} of  $\Lambda$  is also minimal, Laplacian, second-order, pseudo-differential type as desired.
One can obtain the form of $\mathbb{I} $,
$\omega_{\rho}$ and $E$ by comparing \eqref{E} and \eqref{V} with help of the effective covariant derivative \eqref{effective D}, 
\begin{align}\label{identity}
	\tilde{\varphi}_{m}\left(\mathbb{I} \right)^{mn} \tilde{\varphi_{n}}=&\tilde{\varphi}_{m}G^{mn}\tilde{\varphi}_{n},
\end{align}
\begin{align}\label{omega}
	\tilde{\varphi}_{m}\left(\omega_{\rho} \right)^{mn} \tilde{\varphi_{n}}=&\tilde{\varphi}_{m}\left(\frac{1}{2} N_{\rho} \right)^{mn}\tilde{\varphi}_{n},
\end{align}
\begin{equation}\label{P}
	\tilde{\varphi}_{m}E^{mn}\tilde{\varphi}_{n}=\tilde{\varphi}_{m}P^{mn}\tilde{\varphi}_{n}-\tilde{\varphi}_{m}(\omega^\rho)^{mp} {(\omega_\rho)_{p}}^{n}\tilde{\varphi}_{n}-\tilde{\varphi}_{m}(D^\rho\omega_{\rho})^{mn}\tilde{\varphi}_{n}.
\end{equation}
And, the field strength $\Omega_{\alpha \beta} $ associated with the curvature of $\mathcal{D}_{\mu}$ is obtained via
\begin{equation}\label{Omega}
	\tilde{\varphi}_{m}(\Omega_{\alpha \beta})^{mn}\tilde{\varphi}_{n}\equiv \tilde{\varphi}_{m}\left[\mathcal{D}_{\alpha},\mathcal{D}_{\beta} \right]^{mn}\tilde{\varphi}_{n}.
\end{equation} 
The  Seeley-DeWitt coefficients for a manifold without boundary can be determined from the knowledge of $\mathbb{I}$, $E$ and $\Omega_{\alpha \beta}$. In the following, we give expressions for the first three Seeley-DeWitt coefficients $a_0$, $a_2$, $a_4$  in terms of local background invariant parameters \cite{Gilkey:1984xy,Gilkey:1975cd,Vassilevich:2003ll},
%Once $\mathbb{I}$, $E$ and $\Omega_{\alpha \beta}$ got determined then, the first three Seeley De-Witt coefficients $a_0$, $a_2$, $a_4$ for manifold without boundary can be determined from the relation given below where these coefficients are expressed as local background invariant parameters \cite{Gilkey:1984xy,Gilkey:1975cd}
\begin{align}\label{a 2n}
	\begin{split}
		\chi (4\pi)^2 a_0(x)&= \text{tr}~(\mathbb{I}),
		\\
		\chi (4\pi)^2 a_2(x)&=\frac{1}{6}\text{tr}~(6E+R \mathbb{I}),
		\\
		\chi (4\pi)^2 a_4(x)&= \frac{1}{360}\text{tr}~\left\lbrace  60 R E +180 E^2+30 \Omega^{\mu\nu}\Omega_{\mu\nu}+(5R^2-2R^{\mu\nu}R_{\mu\nu}+2R^{\mu\nu\rho\sigma}R_{\mu\nu\rho\sigma})\mathbb{I}\right\rbrace,
	\end{split}
\end{align}
where $R$, $R_{\mu\nu}$ and $R_{\mu\nu\rho\sigma}$ are the usual curvature tensors computed over the background metric. 

%Proceeding the same way Many works/people also studied
There exist many studies about other higher-order coefficients, e.g., $a_6$\cite{Gilkey:1975cd}, $a_8$ \cite{Avramidi,Avramidi2}, $a_{10}$ \cite{VandeVen:1997pf}, and higher heat kernels \cite{Fliegner:1994zc,Fliegner:1997rk}. However, we will limit ourselves up to $a_4$, which helps to determine the logarithmic divergence of the string theoretic black holes. 
%The coefficient $a_4$ determines the logarithmic divergence of the string theoretic black hole \cite{Sen:2012rr}.
Apart from this, the coefficient $a_2$ describes the quadratic divergence and renormalization of Newton constant and $a_0$ encodes the  cosmological constant at one-loop approximation \cite{Castro:2018hsc, Charles:2015nn}.  
%..................................................................................................................................................................
\section{``Non-minimal'' $\mathcal{N}=1$ Einstein-Maxwell supergravity theory}\label{Einstein-Maxwell theory in N eq 1 supergravity}
The gauge action of pure $\mathcal{N}=1$ supergravity, containing spin 2 graviton with  superpartner massless spin 3/2 Rarita-Schwinger fields,  was first constructed and the corresponding supersymmetric transformations were studied by Ferrara et al. \cite{Freedman:1976xh}. Later, this work is extended in \cite{Ferrara} by coupling an additional vector multiplet (matter coupling) to the pure $\mathcal{N}=1$ supergravity. Here the vector multiplet is non-minimally coupled to the supergauged action of $\mathcal{N}=1$ theory.
% First, Ferrara constructed the gauge action of $\mathcal{N}=1$ pure  supergravity, and studied the supersymmetric transformation to the theory of gravitation  \cite{Freedman:1976xh} [56]. 
%The supergauge action of this theory contains gravitational spin 2 field $g_{\mu\nu}$ (graviton) with superpartner massless spin 3/2 Rarita-Schwinger $\psi_{\mu}$ (gravitino) fields. Later, this work is extended to couple it with the matter multiplet, i.e., the vector multiplet in \cite{Ferrara}[46]. The vector multiplet of $\mathcal{N}=1$ theory is non-minimally coupled to the supergauged action.
The resultant \say{non-minimal} $\mathcal{N}=1$ $d=4$ EMSGT is the object of interest of the current paper. Two other special classes of $\mathcal{N}=1$ supergravity theories,  namely \say{minimally coupled} $\mathcal{N}=1$ supergravity theory and  $\mathcal{N}=1$ theory obtained from truncation of $\mathcal{N}=2$ theory, were also studied by Ferrara et al. in \cite{Ferrara:2011qf, Andrianopoli:2001gm}. This \say{non-minimal} $\mathcal{N}=1,d=4$ EMSGT has the same on-shell bosonic and fermionic degrees of freedom. The bosonic sector is controlled by a 4D Einstein-Maxwell theory \cite{Bhattacharyya:2012ss}, while the fermionic sector casts a non-minimal exchange of graviphoton between the gravitino and gaugino species. We aim to calculate the Seeley-DeWitt coefficients in the mentioned theory for the fluctuation of its fields. We also summarize the classical equations of motion
and identities for the theory. 

\subsection{Supergravity solution and classical equations of motion}

The \say{non-minimal} $\mathcal{N}=1$ EMSGT in four dimensions is defined with the following  field contents: a gravitational spin 2 field $g_{\mu\nu}$ (graviton) and an abelian spin 1 gauge field $A_{\mu}$ (photon) along with their superpartners, massless spin 3/2 Rarita-Schwinger $\psi_{\mu}$ (gravitino) and spin 1/2 $\lambda$ (gaugino) fields.
% All the fields are minimally coupled to the gravitational background field. However,
Note that the gaugino field $(\lambda)$  interacts non-minimally with the gravitino field $(\psi_{\mu})$ via non-vanishing  gauge field strength. 

The theory is described by the following action \cite{Ferrara}
%\footnote{We have set $G_N=\frac{1}{16 \pi}$ throughout the paper} 
\begin{equation}\label{full action}
	\mathcal{S}=\int d^4 x \sqrt{g} \mathcal{L}_{\text{EM}}+ \int d^4 x \sqrt{g} \mathcal{L}_{f},
\end{equation}
where, 
%\footnote{The quadratic order fermionic lagrangian $\mathcal{L}_{f}$ is obtained from \cite{Ferrara}.},
\begin{equation}\label{EM action}
	\mathcal{L}_{\text{EM}}=\mathcal{R}-F^{\mu\nu} F_{\mu\nu},
\end{equation}
\begin{equation}\label{L fermions}
	\mathcal{L}_{f}=-\frac{1}{2} \bar{\psi}_{\mu} \gamma^{[\mu\rho\nu]}D_{\rho}\psi_{\nu}-\frac{1}{2}\bar{\lambda} \gamma^{\rho} D_{\rho} \lambda+ \frac{1}{2\sqrt{2}}  \bar{\psi_{\mu}} F_{\alpha \beta} \gamma^{\alpha} \gamma^{\beta} \gamma^{\mu} \lambda.
\end{equation}
$\mathcal{R}$ is the Ricci scalar curvature and $F_{\mu\nu}=\partial_{[\mu}A_{\nu]}$ is the field strength of gauge field $A_{\mu}$.% in the Lagrangian.
% defined by metric $g_{\mu\nu}$ and gauge $A_{\mu}$ fields, respectively. 

%Now, we will derive the equation of motion of $\mathcal{N}=1$ $d=4$ EMSGT.
Let us consider an arbitrary background solution denoted as  $(\bar{g}_{\mu\nu},\bar{A}_{\mu})$ satisfying the classical equations of motion of Einstein-Maxwell theory. The solution is then embedded in $\mathcal{N} = 1$ supersymmetric theory in four dimensions. The geometry of  action \eqref{full action} is defined by the Einstein equation,
% Now we are interested in obtaining the equation of motion of this combined $\mathcal{N}=1$ $d=4$ Einstein-Maxwell supergravity theory obtained after the embedding the solution of Einstein-Maxwell theory to $\mathcal{N}=1$ supersymmetric theory in four dimensions. 
%Now we begin by writing down the Einstein equation which also defines the geometry of action \eqref{full action}
\begin{equation}\label{A}
	R_{\mu\nu}-\frac{1}{2}\bar{g}_{\mu\nu}R=8 \pi G_N T_{\mu\nu}=2 \bar{F}_{\mu\rho} \bar{F_{\nu}}^{\rho}-\frac{1}{2} \bar{g}_{\mu\nu} \bar{F}^{\rho \sigma} \bar{F}_{\rho \sigma},
\end{equation}
where $\bar{F}_{\mu\nu}=\partial_{[\mu}\bar{A}_{\nu]}$ is the  background gauge field strength.
Equation \eqref{A} also defines background curvature invariants in terms of field strength tensor.
%\begin{equation}\label{C}
%R_{\mu\nu}-\frac{1}{2}g_{\mu\nu}R=2 \bar{F}_{\mu\rho}\bar{{F_{\nu}}^{\rho}}-\frac{1}{2}g_{\mu\nu}\bar{F}^{\rho \sigma} \bar{F}_{\rho \sigma}
%\end{equation}
For such a geometrical space-time, the trace of \eqref{A} yields $R=0$ as a classical solution. Hence one can ignore the terms proportional to $R$ in further calculations. The reduced  equation of motion for the $\mathcal{N}=1$, $d=4$ EMSGT is thus given as
\begin{equation}\label{D}
	R_{\mu\nu}=2 \bar{F}_{\mu\rho}\bar{F_{\nu}}^\rho-\frac{1}{2} \bar{g}_{\mu\nu} \bar{F}^{\rho \sigma} \bar{F}_{\rho \sigma}.
\end{equation} 
%Apart from the classical equation of motion \eqref{D},
The theory also satisfies the Maxwell equations and  Bianchi identities of the form,
\begin{equation}\label{MR}
	D^{\mu}\bar{F}_{\mu\nu}=0, \enspace
	D_{[\mu}\bar{F}_{\nu\rho]}=0,\enspace R_{\mu[\nu\theta\phi]}=0.
\end{equation} 
The equations of motion \eqref{D}, \eqref{MR} and  the condition $R=0$ play a crucial role in simplifying the Seeley-DeWitt coefficients for the $\mathcal{N}=1$, $d=4$ EMSGT theory and express them in terms of local background invariants. 
%\subsection{Equation of motion in $\mathcal{N}=1$ EMSGT}
%.................................................................................................................................................................

\section{Heat kernels in ``non-minimal'' $\mathcal{N}=1$, $d=4$ EMSGT}\label{Heat Kernels in N eq 1 EMSGT}

In this  section, we study the spectrum of one-loop action corresponding to the quantum fluctuations in the background fields for the \say{non-minimal} $\mathcal{N}=1$ EMSGT in a compact four dimensional Riemannian manifold without boundary.
We will focus on the computation of the first three Seeley-DeWitt coefficients for the fluctuated bosonic and fermionic fields present in the concerned theory using the heat kernel tool, as described in Section \ref{Heat kernel Technique}. In the later sections, we throw light on the applicability of these computed coefficients from the perspective of logarithmic correction to the entropy of extremal black holes.
%...................................................................................................................................................................................................................................................................................

\subsection{Bosonic sector}\label{bosons}
%...................................................................................................................................................................................................................................................................................
Here we are interested in the analysis  of quadratic order fluctuated action $\mathcal{S}_2$ and computation of the Seeley-DeWitt coefficients for non-supersymmetric 4D Einstein-Maxwell theory with only bosonic fields, i.e., $g_{\mu\nu}$ and $A_{\mu}$, described by the  action \eqref{EM action}. 
We consider the following fluctuations of the fields present in the theory
\begin{equation}\label{bosonic fluctuation}
	g_{\mu\nu}=\bar{g}_{\mu\nu}+\tilde{g}_{\mu\nu},\enspace A_{\mu}=\bar{A}_{\mu}+\tilde{A}_{\mu}.
\end{equation}
The resulting quadratic order fluctuated action $\mathcal{S}_2$ is then studied and the required coefficients were calculated following the general
Seeley-DeWitt technique \cite{Vassilevich:2003ll} for heat kernel.
%  by Sen et al. \cite{Bhattacharyya:2012ss}, also by Larsen et.al \cite{Charles:2015nn}  and reviewed in our recent work \cite{Karan:2019gyn}. 
The first three  Seeley-DeWitt coefficients for bosonic fields read as\footnote{One may go through our earlier work \cite{Karan:2019gyn}  for explicit computation of Seeley-DeWitt coefficients for the bosonic gravity multiplet.} 
%their respective work \cite{Bhattacharyya:2012ss,Charles:2015nn} following Seeley-DeWitt expansion of the heat kernel. We have also reviewed both of these tasks in our recent work \cite{Karan:2019gyn} and calculated the required coefficients for the one-loop effective action following the same approach \footnote{Readers are advised to go through \cite{Karan:2019gyn} where the whole process of analysis of one-loop effective action and computation of Seeley-DeWitt coefficient for bosonic gravity multiplets  is described explicitly in our earlier work.}. Then, we computed these coefficients for bosonic fields which were found to be consistent with the results of \cite{Bhattacharyya:2012ss,Charles:2015nn}, and reads as
%...............................................................................................................................................................................................................................
\begin{align}\label{bosonic part}
	\begin{split}
		(4\pi)^2 {a_{0}^{\text{B}}(x)}&=4,
		\\
		(4\pi)^2 {a_{2}^{\text{B}}(x)}&=6 \bar{F}^{\mu\nu} \bar{F}_{\mu\nu},
		\\
		(4\pi)^2 {a_{4}^{\text{B}}(x)}&=\frac{1}{180}\left (199 R_{\mu\nu\rho\sigma}R^{\mu\nu\rho\sigma}+26R_{\mu\nu}R^{\mu\nu} \right).
	\end{split}
\end{align}
%...............................................................................................................................................................................................................................................................................................................................................
\subsection{Fermionic sector}\label{fermions}
%----------------------------------------------------------------------------------------------.......................................................................................................................................................................................................................................................
In this subsection, we deal with
% the analysis of the  spectrum of effective action and computation of Seeley-DeWitt coefficients for 
the superpartner fermionic fields of  $\mathcal{N}=1$, $d=4$ EMSGT, i.e., $\psi_{\mu}$ and  $\lambda$. As discussed in Section \ref{Einstein-Maxwell theory in N eq 1 supergravity}, these fermionic fields are  non-minimally coupled.
%to the background graviton. 
The quadratic fluctuated  action for these fields is given in \eqref{full action}. Before proceeding further, it is necessary to gauge fix the action for gauge invariance. We achieve that by adding a gauge fixing term $\mathcal{L}_{\text{gf}}$
to the lagrangian \eqref{L fermions}, 
\begin{equation}
	\mathcal{L}_{\text{gf}}=\frac{1}{4}\bar{\psi}_{\mu}\gamma^{\mu}\gamma^{\rho}D_{\rho}\gamma^{\nu}\psi_{\nu}.
\end{equation}
This gauge fixing process cancels the gauge degrees of freedom and induces ghost fields to the theory described by the  following lagrangian,
\begin{equation}\label{L ghost}
	\mathcal{L}_{\text{ghost,$f$}}=\bar{\tilde{b}}\gamma^{\rho}D_{\rho}\tilde{c}+\bar{\tilde{e}}\gamma^{\rho}D_{\rho}\tilde{e}.
\end{equation}
%..........................................................................................................................................................................
In \eqref{L ghost}   $\tilde{b}$, $\tilde{c}$ are Faddeev-Popov ghosts, and $\tilde{e}$ is a ghost associated with the unusual nature of gauge fixing \cite{Banerjee:2010qc}. These ghost fields are bosonic ghost having spin half statistics in ghost lagrangian.
The final quadratic fluctuated gauge fixed action for the fermionic sector, thus obtained along with unphysical ghost fields is expressed as
%...................................................................................................................................................................................................................................
\begin{equation}\label{quadratic fluctuated action}
	\mathcal{S}_{2,\text{F}}=\int d^4 x \sqrt{\bar{g}}\Big(\mathcal{L}_{f}+\mathcal{L}_{\text{gf}}+
	\mathcal{L}_{\text{ghost,$f$}}\Big),
\end{equation}
where we have,
\begin{align}\label{Lf}
	\begin{split}
		\mathcal{L}_{f}+\mathcal{L}_{\text{gf}}&=\frac{1}{4}\bar{\psi_{\mu}}\gamma^{\nu}\gamma^{\rho}\gamma^{\mu}D_{\rho}\psi_{\nu}-\frac{1}{2}\bar{\lambda}\gamma^{\rho}D_{\rho}\lambda
		\\
		&\quad+\frac{1}{4\sqrt{2}}\bar{\psi_{\mu}}\bar{F}_{\alpha\beta}\gamma^{\alpha}\gamma^{\beta}\gamma^{\mu}\lambda-\frac{1}{4\sqrt{2}}\bar{\lambda}\gamma^{\nu}\gamma^{\alpha}\gamma^{\beta}\bar{F}_{\alpha\beta}\psi_{\nu}.
	\end{split}
\end{align}
%.........................................................................................................................................................................
%..........................................................................................................................................................................
Our primary task in this subsection is to compute the Seeley-DeWitt coefficients for action \eqref{quadratic fluctuated action} and then add it to the coefficients obtained for the bosonic part \eqref{bosonic part}. 

%We will deal with the computation for fermionic part and ghost part one by one. 
%\subsubsection{Fermionic contribution without ghost}
Let us begin with the computation of required coefficients for fermionic fields without ghost fields. The corresponding quadratic fluctuated  action can be written as 
%defined by field content $\mathcal{L}_{f}+\mathcal{L}_{gf}$ in \eqref{Lf} takes the form
%..........................................................................................................................................................................
\begin{equation}\label{linear action}
	-\frac{1}{2}\int d^4 x \sqrt{ \bar{g}}\tilde{\varphi}_{m}{O}^{mn}\tilde{\varphi}_{n},
\end{equation} 
%.
where $O$ is the differential operator which can be expressed in Dirac form as\footnote{The spinors in \eqref{Lf} satisfy the Majorana condition. But in order to follow the heat kernel method, we first compute the Seeley-DeWitt coefficients for Dirac spinors and then obtain the same for the Majorana spinors.}
%..........................................................................................................................................................................
\begin{align}\label{operator form}
	\begin{split}
		\tilde{\varphi}_{m}O^{mn}\tilde{\varphi}_{n}&=-\frac{i}{2}\bar{\psi}_{\mu}\gamma^{\nu}\gamma^{\rho}\gamma^{\mu}D_{\rho}\psi_{\nu}
		+i\bar{\lambda} \gamma^{\rho} D_{\rho} \lambda \\
		&\quad-\frac{i}{2\sqrt{2}}\bar{\psi}_{\mu}\bar{F}_{\alpha \beta}\gamma^{\alpha}\gamma^{\beta}\gamma^{\mu}  \lambda  +\frac{i}{2\sqrt{2}}\bar{\lambda}\gamma^{\nu}\gamma^{\alpha}\gamma^{\beta}\bar{F}_{\alpha \beta}\psi_{\nu}.
	\end{split}
\end{align}
Since we consider the Euclidean continuation of $\mathcal{N}=1$ supergravity in 4D Einstein-Maxwell theory, the  gamma matrices are Hermitian. Hence,
the differential operator $O^{mn}$ is linear, Dirac type and Hermitian.
In order to compute the Seeley-DeWitt coefficients following the general technique  described in Section \ref{Heat kernel calculational framework and methodology}, 
it is to be made Laplacian. Following \eqref{L} we have,
%..........................................................................\label{Heat kernel Technique}................................................................................................
\begin{align}\label{unreduced lambda}
	\begin{split}
		\tilde{\varphi}_{m}\Lambda^{mn}\tilde{\varphi}_{n}&= \tilde{\varphi}_{m}\left\lbrace {\left(O \right)^{m}}_p \left(O \right)^{pn\dagger}\right\rbrace \tilde{\varphi}_{n}\\
		&=\bar{\psi}_{\mu}\left\{-\frac{1}{4}\gamma^{\tau}\gamma^{\rho}\gamma^{\mu}\gamma^{\nu}\gamma^{\sigma}\gamma_{\tau}D_{\rho}D_{\sigma}+\frac{1}{8} \gamma^{\alpha}\gamma^{\beta}\gamma^{\mu}\gamma^{\nu}\gamma^{\theta}\gamma^{\phi} \bar{F}_{\alpha \beta} \bar{F}_{\theta \phi} \right\} \psi_{\nu}
		\\
		&\quad+\bar{\lambda}\left\{\frac{1}{8}\gamma^{\tau} \gamma^{\alpha}\gamma^{\beta}\gamma^{\theta}\gamma^{\phi}\gamma_{\tau} \bar{F}_{\alpha \beta} \bar{F}_{\theta \phi} -\gamma^{\rho}\gamma^{\sigma} D_{\rho} D_{\sigma}  \right\} \lambda
		\\
		&\quad+\bar{\psi}_{\mu}\left\{-\frac{1}{4\sqrt{2}}\gamma^{\tau}\gamma^{\rho}\gamma^{\mu}\gamma^{\alpha}\gamma^{\beta}\gamma_{\tau} (D_{\rho} \bar{F}_{\alpha \beta}+\bar{F}_{\alpha \beta} D_{\rho})+\frac{1}{2\sqrt{2}}\gamma^{\alpha}\gamma^{\beta}\gamma^{\mu}\gamma^{\rho}\bar{F}_{\alpha \beta} D_{\rho} \right\} \lambda
		\\
		&\quad+\bar{\lambda}\left\{\frac{1}{4\sqrt{2}}\gamma^{\tau}\gamma^{\alpha}\gamma^{\beta}\gamma^{\nu}\gamma^{\rho}\gamma_{\tau} \bar{F}_{\alpha\beta} D_{\rho}-\frac{1}{2\sqrt{2}}\gamma^{\rho}\gamma^{\nu}\gamma^{\alpha}\gamma^{\beta} (D_{\rho} \bar{F}_{\alpha \beta}+\bar{F}_{\alpha \beta}D_{\rho})  \right\}\psi_{\nu}.
	\end{split}
\end{align}
%..........................................................................................................................................................................
The form of $\Lambda^{mn}$ can further be simplified by using various identities and ignoring terms dependent on the Ricci scalar $R$,
%..........................................................................................................................................................................
%Applying \eqref{identities} in \eqref{unreduced lambda}, and setting $R=0$ the form of $\Lambda$  would further take a simplified form
\begin{align}\label{reduced lambda}
	\begin{split}
		\tilde{\varphi}_{m}\Lambda^{mn}\tilde{\varphi}_{n}&=-\mathbb{I}_{4}\bar{g}^{\mu\nu}\bar{\psi}_{\mu}D^{\rho}D_{\rho} \psi_{\nu}-\mathbb{I}_{4}\bar{\lambda}D^{\rho} D_{\rho} \lambda + \bar{\psi}_{\mu}\Big\{R^{\mu\nu}-\frac{1}{2}\gamma^{\nu}\gamma^{\theta}{R^{\mu}}_{\theta}
		\\
		&\quad+\frac{1}{2}\gamma^{\mu}\gamma^{\theta}{R^{\nu}}_{\theta}-\frac{1}{2}\gamma^{\rho}\gamma^{\sigma}{R^{\mu\nu}}_{\rho\sigma}+\frac{1}{8}\gamma^{\alpha}\gamma^{\beta}\gamma^{\mu}\gamma^{\nu}\gamma^{\theta}\gamma^{\phi}\bar{F}_{\alpha \beta} \bar{F}_{\theta \phi}\Big\}\psi_{\nu}
		\\
		&\quad+\frac{1}{4}\bar{\lambda}\Big\{\left(\gamma^{\phi}\gamma^{\alpha}\gamma^{\beta}\gamma^{\theta}+\gamma^{\theta}\gamma^{\beta}\gamma^{\alpha}\gamma^{\phi} \right)\bar{F}_{\alpha\beta}\bar{F}_{\theta\phi} \Big\} \lambda
		\\
		&\quad-\frac{1}{2\sqrt{2}}\bar{\lambda}\Big\{\gamma^{\rho}\gamma^{\nu}\gamma^{\alpha}\gamma^{\beta} (D_{\rho} \bar{F}_{\alpha \beta}) \Big\}\psi_{\nu}
		\\
		&\quad-\frac{1}{2\sqrt{2}}\bar{\psi}_{\mu}\Big\{\left(\gamma^{\beta}\gamma^{\rho}\gamma^{\mu}\gamma^{\alpha}+\gamma^{\alpha}\gamma^{\mu}\gamma^{\rho}\gamma^{\beta} \right)(D_{\rho} \bar{F}_{\alpha \beta})\Big\}\lambda
		\\
		&\quad-\frac{1}{2\sqrt{2}}\bar{\psi}_{\mu}\Big\{\left(\gamma^{\beta}\gamma^{\rho}\gamma^{\mu}\gamma^{\alpha}+\gamma^{\alpha}\gamma^{\mu}\gamma^{\rho}\gamma^{\beta}-\gamma^{\alpha}\gamma^{\beta}\gamma^{\mu}\gamma^{\rho}\right)\bar{F}_{\alpha \beta}  \Big\}D_{\rho}\lambda
		\\
		&\quad+\frac{1}{2\sqrt{2}}\bar{\lambda}\Big\{\left(\gamma^{\rho} \gamma^{\alpha} \gamma^{\beta} \gamma^{\nu} + \gamma^{\nu} \gamma^{\beta} \gamma^{\alpha} \gamma^{\rho}-\gamma^{\rho} \gamma^{\nu}\gamma^{\alpha} \gamma^{\beta} \right)\bar{F}_{\alpha \beta}  \Big\}D_{\rho} \psi_{\nu}.
	\end{split}
\end{align}
The Laplacian operator $\Lambda$  obtained in \eqref{reduced lambda}
fits in the essential format \eqref{E} required for the heat kernel analysis, thus providing the expressions of $\mathbb{I}$, $N^{\rho}$ and $P$ as
%Thus, $\Lambda$ \eqref{reduced lambda} obtained is the Laplacian operator which is required for the heat kernel analysis that fits in format \eqref{E} from where one can obtain the expressions of $N^{\rho}$ and $P$.
\begin{align}
	\tilde{\varphi}_{m}\mathbb{I}^{mn}\tilde{\varphi}_{n}=\bar{\psi_{\mu}}\mathbb{I}_{4}g^{\mu\nu}\psi_{\nu}+\bar{\lambda} \mathbb{I}_{4}\lambda,
\end{align}
\begin{align}\label{expression of P}
	\begin{split}
		\tilde{\varphi}_{m} P^{mn}\tilde{\varphi}_{n}&=\bar{\psi}_{\mu}\Big\lbrace-R^{\mu\nu}\mathbb{I}_{4}+\frac{1}{2}\gamma^{\nu}\gamma^{\alpha}{R^{\mu}}_{\alpha}
		\\
		&\quad-\frac{1}{2}\gamma^{\mu}\gamma^{\alpha}{R^{\nu}}_{\alpha}+\frac{1}{2}\gamma^{\eta}\gamma^{\rho}{R^{\mu\nu}}_{\eta\rho}-\frac{1}{8}\gamma^{\alpha}\gamma^{\beta}\gamma^{\mu}\gamma^{\nu}\gamma^{\theta}\gamma^{\phi} \bar{F}_{\alpha \beta} \bar{F}_{\theta \phi}\Big\rbrace\psi_{\nu}
		\\
		&\quad-\frac{1}{4}\bar{\lambda}\Big\{\gamma^{\phi} \gamma^{\alpha}\gamma^{\beta}\gamma^{\theta}+\gamma^{\theta}\gamma^{\beta}\gamma^{\alpha}\gamma^{\phi} \Big\} \bar{F}_{\alpha \beta}\bar{F}_{\theta \phi}\lambda
		\\
		&\quad+\frac{1}{2\sqrt{2}}\bar{\psi}_{\mu}\Big\{(\gamma^{\beta}\gamma^{\rho}\gamma^{\mu}\gamma^{\alpha}+\gamma^{\alpha}\gamma^{\mu}\gamma^{\rho}\gamma^{\beta})D_{\rho} \bar{F}_{\alpha \beta} \Big\}\lambda
		\\
		&\quad+\frac{1}{2\sqrt{2}}\bar{\lambda}\Big\{\gamma^{\rho}\gamma^{\nu}\gamma^{\alpha}\gamma^{\beta}D_{\rho} \bar{F}_{\alpha \beta} \Big\}\psi_{\nu},
	\end{split}
\end{align}
\begin{align}\label{Nmn}
	\begin{split}
		\tilde{\varphi}_{m}\left(N^{\rho} \right)^{mn} \tilde{\varphi}_{n}&=\frac{1}{2\sqrt{2}}\bar{\psi}_{\mu}\Big\{\left(\gamma^{\beta}\gamma^{\rho}\gamma^{\mu}\gamma^{\alpha}+\gamma^{\alpha}\gamma^{\mu}\gamma^{\rho}\gamma^{\beta}-\gamma^{\alpha}\gamma^{\beta}\gamma^{\mu}\gamma^{\rho} \right) \bar{F}_{\alpha\beta} \Big\}  \lambda
		\\
		&\quad-\frac{1}{2\sqrt{2}}\bar{\lambda}\Big\{\left(\gamma^{\rho}\gamma^{\alpha}\gamma^{\beta}\gamma^{\nu}+\gamma^{\nu}\gamma^{\beta}\gamma^{\alpha}\gamma^{\rho}-\gamma^{\rho}\gamma^{\nu}\gamma^{\alpha}\gamma^{\beta} \right)\bar{F}_{\alpha \beta} \Big\}\psi_{\nu}.
	\end{split}
\end{align}
One can extract $\omega^{\rho}$  from $N^{\rho}$ by using \eqref{omega}, which on further simplification using gamma matrices properties, reads as
\begin{align}\label{omega 2}
	\begin{split}
		\tilde{\varphi}_{m}(\omega^{\rho})^{mn}\tilde{\varphi}_{n}&=\bar{\psi}_{\mu}\left\{-\frac{1}{4\sqrt{2}}\gamma^{\theta} \gamma^{\phi} \gamma^{\rho}\gamma^{\mu}\bar{F}_{\theta \phi}+\frac{1}{\sqrt{2}}\gamma^{\tau} \gamma^{\rho} \bar{{F^{\mu}}}_{\tau}-\frac{1}{\sqrt{2}}\gamma^{\tau}\gamma^{\mu}{\bar{F^{\rho}}_{\tau}} \right\}\lambda
		\\
		&\quad+\bar{\lambda}\left\{ \frac{1}{4\sqrt{2}}\gamma^\nu \gamma^{\rho} \gamma^{\theta} \gamma^{\phi} \bar{F}_{\theta \phi}+\frac{1}{\sqrt{2}} \gamma^{\rho} \gamma^{\tau} \bar{{F^\nu}}_\tau-\frac{1}{\sqrt{2}}\gamma^\nu \gamma^\tau \bar{F^\rho}_{\tau} \right\}\psi_{\nu}.
	\end{split}
\end{align}
The strategy is to express the kinetic differential operator $\Lambda$ in the form that fits in the prescription \eqref{V} and find the required quantities. 
Determination of $E$ and $\Omega_{\alpha \beta}$ from  $P$, $N^{\rho}$ and $\omega^{\rho}$  is now  straightforward by using \eqref{P}, \eqref{Omega},  \eqref{expression of P} and \eqref{omega 2}.\footnote{The full expression of $E$ and $\Omega_{\alpha \beta}$ are given in  \eqref{E Appendix} and \eqref{Omega Appendix}.} One can then calculate the required trace values from the expression of $\mathbb{I}$, $E$ and $\Omega_{\alpha \beta}$. The results are 
% $E$ by determining $\omega^{\rho}$, $N^{\rho}$ and $P$. One can extract $\omega^{\rho}$  from $N^{\rho}$ by using \eqref{omega} which on further simplification using gamma matrices properties reads as
%\begin{align}\label{omega 2}
%\tilde{\varphi}_{m}\left(\omega^{\rho} \right)^{mn} \tilde{\varphi}_{n}=&\frac{1}{4\sqrt{2}}\bar{\psi}_{\mu}\Big(\left(\gamma^{\beta}\gamma^{\rho}\gamma^{\mu}\gamma^{\alpha}+\gamma^{\alpha}\gamma^{\mu}\gamma^{\rho}\gamma^{\beta}-\gamma^{\alpha}\gamma^{\beta}\gamma^{\mu}\gamma^{\rho} \right) \bar{F}_{\alpha\beta} \Big)  \lambda
%\nonumber\\
%&-\frac{1}{4\sqrt{2}}\bar{\lambda}\Big(\left(\gamma^{\rho}\gamma^{\alpha}\gamma^{\beta}\gamma^{\nu}+\gamma^{\nu}\gamma^{\beta}\gamma^{\alpha}\gamma^{\rho}-\gamma^{\rho}\gamma^{\nu}\gamma^{\alpha}\gamma^{\beta} \right)\bar{F}_{\alpha \beta} \Big)\psi_{\nu}.
%\end{align}
\begin{align}\label{Trace}
	\begin{split}
		\text{tr}\thickspace\left(  \mathbb{I} \right) &=16+4=20,
		\\
		\text{tr} \thickspace \left( E  \right) &=-8 \bar{F}^{\mu \nu}\bar{F}_{\mu\nu},
		\\
		\text{tr}\thickspace \left( E^2 \right) &=10(\bar{F}^{\mu\nu}\bar{F}_{\mu\nu})^2+3R^{\mu\nu}R_{\mu\nu}+2R^{\mu\nu\theta\phi}R_{\mu\nu\theta\phi}-2R_{\alpha\beta\theta\phi}\bar{F}^{\alpha\beta}\bar{F}^{\theta\phi},
		\\
		\text{tr} \thickspace \left( \Omega^{\alpha \beta}\Omega_{\alpha\beta} \right) &=-\frac{13}{2}R^{\mu\nu\theta\phi}R_{\mu\nu\theta\phi}+12R_{\alpha\beta\theta\phi}\bar{F}^{\alpha\beta}\bar{F}^{\theta\phi}-6R^{\mu\nu}R_{\mu\nu}
		\\
		&\quad-60(\bar{F}^{\mu\nu}\bar{F}_{\mu\nu})^2.
	\end{split}
\end{align}
In the above trace result of $\mathbb{I}$, the values 16 and 4 correspond to the degrees of freedom associated with the gravitino and gaugino fields, respectively.
%In the above computation of trace of identity matrix, 16 and 4  corresponds to the degrees of freedom  associated with spin 3/2 Rarita Schwinger field and spin 1/2 Dirac field, respectively.
So, the relevant Seeley-DeWitt coefficients for the gauged fermionic part, computed using \eqref{Trace} in  \eqref{a 2n} for the Majorana fermions, are
\begin{align}\label{fermion coefficients}
	\begin{split}
		(4\pi)^2 a_{0}^{f}(x)&=-10,
		\\
		(4\pi)^2 a_{2}^{f}(x)&=4 \bar{F}^{\mu\nu} \bar{F}_{\mu\nu},
		\\
		(4\pi)^2a_{4}^{f}(x)&=-\frac{1}{144}\left(41 R^{\mu\nu\theta\phi} R_{\mu\nu\theta\phi}+64R^{\mu\nu}R_{\mu\nu} \right).
	\end{split}
\end{align}
Here a -ve sign corresponding to the value of $\chi$ from \eqref{a 2n} for fermion spin-statistics and a factor of 1/2 for considering Majorana degree of freedom are imposed manually in determining the above coefficients.

%\subsubsection{Ghost contribution}
Next, we aim to  compute the Seeley-DeWitt coefficients for the ghost fields described by the  lagrangian \eqref{L ghost}.  The ghost fields 
%\begin{equation}
%\mathcal{L}_{ghost}=b \gamma^{\mu}D_{\mu}c+e\gamma^{\mu}D_{\mu}e
%\end{equation}
$\tilde{b},\tilde{c}$ and $\tilde{e}$ are three minimally coupled Majorana  spin $1/2$ fermions. So, the contribution of Seeley-DeWitt coefficients from the ghost fields will be -3 times of the coefficients of a free Majorana spin $1/2$ field,
\begin{equation}\label{ghost contribution}
	a_{2n}^{\text{ghost},f}(x)=-3 a_{2n}^{1/2,f}(x),
\end{equation}
where $a_{2n}^{1/2,f}(x)$ is the Seeley-DeWitt coefficients for a free Majorana spin 1/2 field calculated in \cite{Karan:2017txu}. 
%The computation of required coefficients for spin (1/2) fields has already been done in our earlier work in .
The Seeley-DeWitt coefficients for ghost sector are
\begin{align}\label{ghost contribution for a0 to a4}
	\begin{split}
		(4\pi)^2 a_{0}^{\text{ghost},f}(x)&=6,
		\\
		(4\pi)^2 a_{2}^{\text{ghost},f}(x)&=0,
		\\
		(4\pi)^2 a_{4}^{\text{ghost},f}(x)&=-\frac{1}{240}\left(7 R^{\mu\nu\theta \phi} R_{\mu\nu\theta\phi} + 8R_{\mu\nu}R^{\mu\nu} \right).
	\end{split}
\end{align}
The net Seeley-DeWitt coefficients for the fermionic sector can be computed by summing up the results \eqref{fermion coefficients} and \eqref{ghost contribution for a0 to a4},
%Following the above relation \eqref{Net fermionic contribution}, and using \eqref{fermion coefficients} together with \eqref{ghost contribution for a0 to a4},
%the final Seeley-DeWitt coefficient for the fermionic multiplet will be
\begin{align}\label{fermion part}
	\begin{split}
		(4\pi)^2 a_{0}^{\text{F}}(x)&=-4,
		\\
		(4\pi)^2 a_{2}^{\text{F}}(x)&=4\bar{F}^{\mu\nu}\bar{F}_{\mu\nu},
		\\
		(4\pi)^2 a_{4}^{\text{F}}(x)&=-\frac{1}{360}\left(113 R^{\mu\nu\theta\phi} R_{\mu\nu\theta\phi}+172 R^{\mu\nu}R_{\mu\nu} \right).
	\end{split}
\end{align} 
%..................................................................................................................................................................................................................................
\subsection{Total Seeley-DeWitt coefficients}\label{total seeley-Witt coefficients}
%.........................................................................................................................................................................
The total Seeley-DeWitt Coefficients obtained by adding the bosonic contribution \eqref{bosonic part} and the fermionic contribution \eqref{fermion part} would be, 
\begin{align}\label{total a2n}
	\begin{split}
		(4\pi)^2 a_{0}^{\text{B+F}}(x)&=0,
		\\
		(4\pi)^{2} a_{2}^{\text{B+F}}(x)&=10 \bar{F}^{\mu\nu}\bar{F}_{\mu\nu},
		\\
		(4\pi)^2 a_{4}^{\text{B+F}}(x)&=\frac{1}{24}\left(19 R^{\mu\nu\theta\phi}R_{\mu\nu\theta\phi}-8R^{\mu\nu}R_{\mu\nu} \right).
	\end{split}
\end{align} 
%The coefficients $a_{0}^{B+F}$ vanishes due to equal contribution of degrees of freedom for bosonic and fermionic matter multiplet and
The coefficient $a_4^{\text{B+F}}$ obtained in \eqref{total a2n} is independent of the terms $R_{\mu\nu\theta\phi} \bar{F}^{\mu\nu} \bar{F}^{\theta\phi}$ and $(\bar{F}^{\mu\nu} \bar{F}_{\mu\nu})^{2}$, and only depends upon the background metric which predicts the rotational invariance of the result under electric and magnetic duality.
%..................................................................................................................................................................................................................................
\section{Logarithmic correction of extremal black holes in ``non-minimal'' $\mathcal{N}=1,d=4$ EMSGT}\label{Application of Seeley-DeWitt coefficients in Logarithmic correction of extremal Black holes}
In this section, we discuss the applicability of Seeley-DeWitt coefficients in the area of black hole physics. We use our results, particularly $a_{4}$ coefficient for the computation of logarithmic correction part of extremal black hole entropy in \say{non-minimal} $\mathcal{N}=1,d=4$ EMSGT.

\subsection{General methodology for computing logarithmic correction using Seeley-DeWitt coefficient}\label{General methodology for computing Logarithmic corrected entropy of black holes}
We review the particular procedure adopted in \cite{Bhattacharyya:2012ss,Karan:2019gyn,
	Banerjee:2010qc,Banerjee:2011jp,Sen:2012rr,Sen:2011ba} for determining the logarithmic correction of entropy of extremal black holes using quantum entropy function formalism \cite{Sen:2008yk,Sen:2008vm,Sen:2009vz}. An extremal black hole is defined with near horizon geometry $AdS_2 \times \mathcal{K}$, where $\mathcal{K}$ is a compactified space fibered over $AdS_2$ space. The quantum corrected entropy ($S_{\text{BH}}$) of an extremal black hole is related to the near horizon partition function ($ \mathcal{Z}_{AdS_2}$) via $AdS_2/CFT_1$ correspondence \cite{Sen:2008yk} as
\begin{equation}\label{x}
	e^{S_{\text{BH}}-E_0\beta} = \mathcal{Z}_{AdS_2},
\end{equation}
where $E_0$ is ground state energy of the extremal black hole carrying charges and $\beta$ is the boundary length of regularized $AdS_2$. We denote coordinate of $AdS_2$ by $(\eta, \theta)$ and that of $\mathcal{K}$ by $(y)$. If the volume of $AdS_2$ is made finite by inserting an IR cut-off $\eta \leq \eta_0$ for regularization, then the one-loop correction to the partition function $ \mathcal{Z}_{AdS_2}$ is presented as
\begin{align}
	\begin{split}
		\mathcal{Z}_{AdS_2}^{\text{1-loop}} &= e^{-W},
	\end{split}
\end{align}
with the one-loop effective action,
\begin{align}
	\begin{split}
		W &= \int_0^{\eta_0}  d\eta \int_{0}^{2\pi}d\theta \int_{\text{hor}} d^2 y \sqrt{\bar{g}}\Delta\mathcal{L}_{\text{eff}}\\
		&= 2 \pi  (\cosh \eta_{0}-1)\int_{\text{hor}} d^2 y  G(y)\Delta \mathcal{L}_{\text{eff}},
	\end{split}
\end{align}
where $G(y)= \sqrt{\bar{g}}/\sinh\eta$ and $\Delta \mathcal{L}_{\text{eff}}$ is the lagrangian density corresponding to the one-loop effective action. The integration limit over $(y)$ depends upon the geometry associated with $\mathcal{K}$ space. The above form of $W$ can be equivalently interpreted as \cite{Sen:2012rr,Sen:2011ba}
\begin{equation}\label{y}
	W = -\beta \Delta E_{0}+\mathcal{O}(\beta^{-1})- 2\pi \int_{\text{hor}} d^2 y  G(y)  \Delta \mathcal{L}_{\text{eff}},
\end{equation}	
where $\beta = 2\pi\sinh \eta_{0}$ and the shift in the ground state energy $\Delta E_{0} = -\int d^2 y  G(y) \Delta \mathcal{L}_{\text{eff}}$. For the extremal black holes, the $W$ form \eqref{y} extracts the one-loop corrected entropy ($\Delta S_{\text{BH}}$) from the $\beta$ independent part of the relation \eqref{x} as
\begin{equation}
	\Delta S_{\text{BH}}= 2\pi \int_{\text{hor}}d^2 y  G(y) \Delta \mathcal{L}_{\text{eff}}.
\end{equation}
These one-loop corrections serve as logarithmic corrections to the entropy of extremal black holes if one considers only massless fluctuations in the one-loop. In the standard heat kernel analysis of the one-loop effective lagrangian $\Delta \mathcal{L}_{\text{eff}}$, only the $a_4(x)$ Seeley-DeWitt coefficient provides a term proportional to logarithmic of horizon area $A_H$ \cite{ Karan:2019gyn, Bhattacharyya:2012ss},
\begin{equation}
	\Delta \mathcal{L}_{\text{eff}} \simeq -\frac{\chi}{2}\Big(a_4(x)-(Y-1)K^{\text{zm}}(x,x;0) \Big) \ln A_{H},
\end{equation}
where $K^{\text{zm}}(x,x;0)$ is the heat kernel contribution for the modes of $\Lambda$ having vanishing eigenvalues and $Y$ is the scaling dimension associated with zero mode corrections of massless fields under considerations. For extremal black holes defined in large charge and angular momentum limit, the present set up computes the logarithmic correction into two parts -- a local part ($\Delta S_{\text{local}}$) controlled by the $a_4(x)$ coefficient and a zero mode part ($\Delta S_{\text{zm}}$) controlled by the zero modes of the massless fluctuations in the near-horizon,\footnote{Total logarithmic correction is obtained as $\Delta S_{\text{BH}} = \Delta S_{\text{local}} + \Delta S_{\text{zm}}$.}  
\begin{align}
	\Delta S_{\text{local}} &= -\pi \int_{\text{hor}} d^2 y  G(y)\chi a_{4}(x) \ln A_{H}, \label{Slocal}\\
	\Delta S_{\text{zm}} &= -\frac{1}{2}\bigg(\sum_{r\in \text{B}}(Y_{r}-1)M^{r}-\sum_{r\in \text{F}}(2Y_{r}-1)M^{r}\bigg) \ln A_{H} \label{Snzmzm}.
\end{align}
$M_r$ and $Y_r$ are respectively the number of zero modes and scaling dimension for different boson (B) and fermion (F) fluctuations. These quantities take different values depending on dimensionality and types of field present. In four dimensions we have \cite{Bhattacharyya:2012ss,Charles:2015nn,Castro:2018hsc,Banerjee:2011jp,Sen:2012rr,Sen:2011ba,Sen:2012dw,Banerjee:2010qc,Keeler:2014bra},
\begin{align}
	Y_{r} &=
	\begin{cases} \label{Y}
		2 \>\>\>\>\>: \text{metric},\\
		1 \>\>\>\>\>: \text{gauge fields},\\
		1/2\> : \text{spin 1/2 fields},\\
		3/2\> : \text{gravitino fields},
	\end{cases}
	\\
	M_{r} &=
	\begin{cases}\label{X}
		3+X : \text{metric in extremal limit},\\
		0 \hspace{0.35in}  : \text{gravitino field for non-BPS solutions},
	\end{cases}
\end{align}
where $X$ is the number of rotational isometries associated with the angular momentum $J$ of black holes,
%where  is the number of zero mode of the rotational isometry associated with the angular momentum  of black holes,
\begin{equation}\label{Z}
	X=
	\begin{cases} 1 \>\>\>\>\forall J_{3}=\text{constant},J^2= \text{arbitrary},\\
		3 \>\>\>\>\forall J_{3}=J^2=0.
	\end{cases}
\end{equation}  
One should not be concerned about the $M_r$ values of the gauge and gaugino fields because their particular $Y_r$ values suggest that they have no zero-mode correction contribution in the formula \eqref{Snzmzm}.
The local mode contribution \eqref{Slocal} for the extremal solutions of the  $\mathcal{N}=1,d=4$ EMSGT  can be computed from the coefficient $a_4(x)$ \eqref{total a2n} in the near-horizon geometry.

\subsection{ Extremal black holes in ``non-minimal'' $\mathcal{N}=1,d=4$ EMSGT and their logarithmic corrections}

An $\mathcal{N}=1,d=4$ EMSGT can have Kerr-Newman, Kerr and Reissner-Nordstr\"{o}m solutions. The metric $\bar{g} $ of general Kerr-Newman black hole with mass $M$, charge $Q$ and angular momentum $J$ is given by 
\allowdisplaybreaks{
	\begin{align}\label{Kerr newman}
		\begin{split}
			ds^2
			&=-\frac{r^2+b^2 \cos^{2}\psi-2Mr+Q^2}{r^2+b^2 \cos^{2}\psi}dt^2
			\\
			&\quad+\frac{r^2+b^2\cos^2\psi}{r^2+b^2-2Mr+Q^2}dr^{2}+(r^2+b^2\cos^{2}\psi)d\psi^{2}
			\\
			&\quad+\left(\frac{(r^2+b^2 \cos^2\psi)(r^2+b^2)+(2Mr-Q^2)b^2 \sin^2\psi}{r^2+b^2 \cos^2 \psi} \right)\sin^2\psi d\phi^{2}
			\\
			&\quad+2\frac{(Q^2-2Mr)b}{r^2+b^2\cos^2\psi}\sin^{2}\psi dt d\phi.
		\end{split}
\end{align}}
where $b= J/M$. The horizon radius is given as
\begin{equation}
	r=M+\sqrt{M^2-(Q^2+b^2)}.
\end{equation}
At the extremal limit $M \to \sqrt{b^2+Q^2}$, the classical entropy of the black hole in near horizon becomes
\begin{equation}
	S_{\text{cl}}=16 \pi^{2} (2b^2+Q^2),
\end{equation}
and the extremal near horizon Kerr-Newman metric is given by \cite{Bhattacharyya:2012ss} 
\begin{align}
	\begin{split}
		ds^2&=(Q^2+b^2+b^2\cos^{2}\psi)(d\eta^2+\sinh^2\eta d\theta^{2}+d\psi^2)
		\\
		&\quad+\frac{(Q^2+2b^2)^2}{(Q^2+b^2+b^2\cos^{2}\psi)} \sin^2\psi \left(d\chi
		-i\frac{2Mb}{Q^2+2b^2}(\cosh \eta-1)d\theta \right)^2.
	\end{split}
\end{align}
For the purpose of using the $a_4(x)$ result \eqref{total a2n} in the relation \eqref{Slocal}, we consider the background invariants \cite{Henry:2000wd,Cherubini:2002we},
\begin{align}\label{K}
	\begin{split}
		R^{\mu\nu\rho\sigma}R_{\mu\nu\rho\sigma}&=\frac{8}{(r^2+b^2 \cos^2 \psi)^6}\bigg\lbrace 6M^2(r^{6}-15b^2r^4 \cos^2 \psi+15 b^4 r^2 \cos^4 \psi 
		\\
		& \quad-b^6 \cos^6 \psi)-12 M Q^2 r(r^4-10 r^2 b^2 \cos^2 \psi +5b^4 \cos^4 \psi)
		\\
		& \quad +Q^4(7r^4-34 r^2 b^2 \cos^2 \psi +7 b^4 \cos^4 \psi) \bigg\rbrace,
		\\
		R^{\mu\nu}R_{\mu\nu}&=\frac{4Q^4}{(r^2+b^2 \cos^2 \psi)^4},
		%\label{R},
		\\
		{G(y)}_{\text{KN}} &= G(\psi)= (Q^2+b^2+b^2\cos^2\psi)(Q^2+2 b^2)\sin\psi.
	\end{split}
\end{align}
In the  extremal limit, one can find \cite{Bhattacharyya:2012ss}
\begin{align}
	\int_{\text{hor}} d \psi d\phi G(\psi) R^{\mu\nu\rho\sigma} R_{\mu\nu\rho\sigma}&=\frac{8\pi}{e (e^2+1)^{5/2}(2e^2+1)}
	\bigg\lbrace 3 (2e^2+1)^2 \tan^{-1}\left(\frac{e}{\sqrt{e^2+1}}\right) \nonumber\\
	&\qquad +e\sqrt{e^2+1}(-8e^6-20e^4-8e^2+1)\bigg\rbrace,\label{int k}\\
	\int_{\text{hor}} d\psi d\phi G(\psi) R^{\mu\nu}R_{\mu\nu}&=\frac{2 \pi}{e (e^2+1)^{5/2}(2e^2+1)}
	\bigg\lbrace3(2e^2+1)^2\tan^{-1}\left(\frac{e}{\sqrt{e^2+1}}\right)
	\nonumber\\
	&\qquad+e \sqrt{e^2+1}(8e^2+5)\bigg\rbrace,\label{int R}
\end{align}
where $e=b/Q$. 
The local mode contribution  in the logarithmic correction to the  entropy of Kerr-Newman extremal black holes in $\mathcal{N}=1,d=4$ EMSGT can be drawn by  using the results (\ref{total a2n}), (\ref{int k})  and (\ref{int R}) in the relation \eqref{Slocal},
\begin{align}\label{KNnzm}
	\begin{split}
		\Delta S_{\text{local,KN}}&=\frac{1}{48}\bigg\lbrace -\frac{51 (2e^2+1)\tan^{-1}\left(\frac{e}{\sqrt{e^2+1}}\right)}{e(e^2+1)^{5/2}}\\
		& \quad+\frac{(152 e^6+380 e^4 +168 e^2 - 9)}{(e^2+1)^2(2e^2+1)}\bigg\rbrace\ln A_{H}.
	\end{split} 
\end{align}
For Kerr black holes we set the limit  $e\to \infty$ in (\ref{KNnzm}), which gives the local mode contribution,
%..........................................................................................................................................................................
\begin{equation}\label{Knzm}
	\Delta S_{\text{local,Kerr}}=\frac{19}{12}\ln A_{H}.
\end{equation}
And for Reissner-Nordstr\"{o}m black holes, we set the limit  $e \to 0$ in (\ref{KNnzm}), yielding the following result
\begin{equation}\label{RNnzm}
	\Delta S_{\text{local,RN}}=-\frac{5}{4}\ln A_{H}.
\end{equation}
The extremal black holes in $\mathcal{N}=1,d=4$ Einstein-Maxwell supergravity theory are  non-BPS in nature\cite{Andrianopoli:2007rm,Ferrara:2011qf}.
Hence, the zero mode contribution to  the correction in entropy of non-BPS extremal Kerr-Newman, Kerr and Reissner-Nordstr\"{o}m black holes are computed by using \eqref{X} to \eqref{Z} in the relation \eqref{Snzmzm} as
\begin{align}
	\Delta S_{\text{zm,KN}} &=-\frac{1}{2}(2-1)(3+1)\ln A_{H}=-2\ln A_{H},\label{KNzm}\\
	\Delta S_{\text{zm,Kerr}} &=-\frac{1}{2}(2-1)(3+1)\ln A_{H}=-2\ln A_{H},\label{Kzm}\\
	\Delta S_{\text{zm,RN}} &=-\frac{1}{2}(2-1)(3+3)\ln A_{H}=-3\ln A_{H}.\label{RNzm}
\end{align}
Merging the above local and zero mode contributions, the total logarithmic correction $\Delta S_{BH}$ ($ = \Delta S_{\text{local}} + \Delta S_{\text{zm}}$) to the entropy of extremal Kerr-Newman, Kerr and Reissner-Nordstr\"{o}m black holes in \say{non-minimal} $\mathcal{N}=1,d=4$ EMSGT are computed as
\begin{align}
	\Delta S_{\text{BH,KN}}&=\frac{1}{48}\bigg\lbrace -\frac{51 (2e^2+1)\tan^{-1}\left(\frac{e}{\sqrt{e^2+1}}\right)}{e(e^2+1)^{5/2}}\nonumber\\
	& \quad+\frac{(152 e^6+380 e^4 +168 e^2 - 9)}{(e^2+1)^2(2e^2+1)}\bigg\rbrace\ln A_{H}-2\ln A_{H},\label{KNFinal}\\
	\Delta S_{\text{BH,Kerr}} &=\frac{19}{12}\ln A_{H}-2\ln A_{H}=-\frac{5}{12}\ln A_{H},\label{KFinal}\\
	\Delta S_{\text{BH,RN}} &=-\frac{5}{4}\ln A_{H}-3\ln A_{H}=-\frac{17}{4} \ln A_{H}.\label{RNFinal}
\end{align}

\section{Concluding remarks} \label{summary}
In summary, we have analyzed the trace anomalies of quadratic fluctuated fields in the \say{non-minimal} $\mathcal{N}=1,d=4$ EMSGT and calculated the first three Seeley-DeWitt coefficients using the approach described in  Section \ref{Heat kernel calculational framework and methodology}. An essential advantage of the Seeley-DeWitt coefficient results \eqref{total a2n} is that these are not background-geometry specific and hence can be used for any arbitrary solution within the \say{non-minimal} $\mathcal{N}=1,d=4$ EMSGT. 

As an application, we have used the third coefficient $a_4$ in the quantum entropy function formalism and computed the local contribution of logarithmic correction to the Bekenstein-Hawking entropy of extremal black holes in the \say{non-minimal}$\mathcal{N}=1,d=4$ EMSGT. The zero mode contribution of the logarithmic correction to the entropy  has been computed separately by analyzing the number of zero modes and the scaling dimensions of the fields in the theory. We, therefore, obtain the net logarithmic correction to the Bekenstein-Hawking entropy of extremal Kerr-Newman \eqref{KNFinal}, Kerr \eqref{KFinal} and Reissner-Nordstr\"{o}m  \eqref{RNFinal} non-BPS black holes in the \say{non-minimal} $\mathcal{N}=1,d=4$ EMSGT. These results are new and significant.  In contrast, the logarithmic correction to the entropy of extremal Reissner-Nordstr\"{o}m black holes in two other classes of $\mathcal{N}=1,d=4$ EMSGT are computed in \cite{Ferrara:2011qf}. However, our work deals with a more general $\mathcal{N}=1,d=4$ EMSGT theory where the vector multiplet is non-minimally coupled the supergravity multiplet. Due to the presence of non-minimal coupling, it is impossible to reproduce the results of \cite{Ferrara:2011qf} from our work and vice-versa. It will also be interesting to reproduce these results by other methods of the heat kernel. The calculated results may also give some valuable inputs for developing insight regarding $\mathcal{N}=1$ Einstein-Maxwell supergravity theory in four dimensions and can also provide an infrared window to any microscopic theory for the same.

\section*{Acknowledgments}
The authors would like to thank Ashoke Sen and Rajesh Kumar Gupta for valuable discussions and useful comments on the work. We also acknowledge IISER-Bhopal for the hospitality during the course of this work.

\appendix
\section{Trace calculations}
\label{Appendix: E and omega}
%\begin{ceqn}
Here, we will explicitly show the expression of $E$ and $\Omega_{\alpha \beta}$, and the computation of traces of $E$, $E^2$ and $\Omega_{\alpha \beta}\Omega^{\alpha \beta}$ for the fermionic sector of $\mathcal{N}=1$, $d=4$ EMSGT. The trace results will be used to calculate $a_4$ required in the logarithmic correction to the entropy of extremal black holes.
In trace calculations, we have used the following identities \cite{Karan:2019gyn}
\begin{equation}
	\begin{aligned}
		&(D_{\rho}\bar{F}_{\mu\nu})(D^{\rho}\bar{F}^{\mu\nu})=R_{\mu\nu\rho\sigma}\bar{F}^{\mu\nu}\bar{F}^{\rho \sigma}-R^{\mu\nu}R_{\mu\nu},
		\\
		&(D_{\mu}\bar{F_{\rho}}^{\nu})(D_{\nu}\bar{F}^{\rho \mu})=\frac{1}{2}(R_{\mu\nu\rho\sigma}\bar{F}^{\mu\nu}\bar{F}^{\rho \sigma}-R^{\mu\nu}R_{\mu\nu}).
	\end{aligned}
\end{equation}
The expression of $E$ by using $\omega_{\rho}$ \eqref{omega 2} will be
\begin{align}
	\begin{split}\label{E Appendix}
		\tilde{\varphi}_{m}E^{mn}\tilde{\varphi}_{n}=& \bar{\psi}_{\mu}E^{\psi_{\mu}\psi_{\nu}}\psi_{\nu}+\bar{\lambda} E^{\lambda \lambda} \lambda+\bar{\psi}_{\mu}E^{\psi_{\mu} \lambda} \lambda+\bar{\lambda} E^{\lambda \psi_{\nu}}\psi_{\nu},
		\\
		=&\bar{\psi}_{\mu}\Big\{-\frac{3}{8}\gamma^{\alpha}\gamma^{\beta}\gamma^{\mu}\gamma^{\nu}\gamma^{\theta}\gamma^{\phi}\bar{F}_{\alpha \beta}\bar{F}_{\theta \phi} -\frac{1}{4}\gamma^{\theta}\gamma^{\phi}\gamma^{\alpha}\gamma^{\mu}\bar{F}_{\theta \phi} \bar{F^\nu}_\alpha
		\\
		&-\frac{1}{4}\gamma^{\theta}\gamma^{\phi}\gamma^{\alpha}\gamma^{\nu}\bar{F}_{\theta \phi} \bar{F^\mu}_\alpha+\frac{1}{8}\bar{g}^{\mu \nu}\gamma^{\alpha}\gamma^{\beta}\gamma^{\theta}\gamma^{\phi}\bar{F}_{\alpha \beta}\bar{F}_{\theta \phi}-\frac{3}{2} \gamma^{\theta}\gamma^{\phi}\bar{F}^{\mu \nu}\bar{F}_{\theta \phi}
		\\
		&-3\gamma^{\alpha}\gamma^{\beta}\bar{F^\mu}_\alpha\bar{F^\nu}_\beta+\frac{1}{2}\gamma^{\mu}\gamma^{\alpha} {R^\nu}_\alpha-\gamma^{\nu}\gamma^{\alpha}{R^\mu}_\alpha+\frac{1}{2}\gamma^{\eta}\gamma^{\rho}{R^{\mu\nu}}_{\eta\rho}
		\\
		& +\frac{3}{4}\gamma^{\mu}\gamma^{\nu} \bar{F}^{\theta\phi}\bar{F}_{\theta\phi}-\frac{1}{2}\bar{g}^{\mu\nu}\mathbb{I}_{4}\bar{F}^{\theta\phi}\bar{F}_{\theta\phi}+\mathbb{I}_{4}R^{\mu\nu}\Big\}\psi_{\nu}+\bar{\lambda}\Big\{\frac{1}{2}\gamma^{\alpha}\gamma^{\beta}\gamma^{\theta}\gamma^{\phi}\bar{F}_{\alpha \beta} \bar{F}_{\theta\phi} \Big\}\lambda 
		\\
		&+\bar{\psi}_{\mu}\Big\{-\frac{3}{4\sqrt{2}} \gamma^{\theta} \gamma^{\phi} \gamma^{\rho}\gamma^{\mu} D_{\rho} \bar{F}_{\theta \phi}+\frac{1}{\sqrt{2}}\gamma^\phi \gamma^ {\rho} D_{\rho} \bar{F^\mu}_\phi+\frac{1}{\sqrt{2}} \gamma^{\theta} \gamma^\phi D^{\mu} \bar{F}_{\theta \phi} \Big\} \lambda
		\\
		&+\bar{\lambda}\Big\{-\frac{3}{4\sqrt{2}}\gamma^{\nu} \gamma^{\rho} \gamma^{\theta} \gamma^{\phi} D_{\rho} \bar{F}_{\theta \phi}-\frac{1}{\sqrt{2}}\gamma^{\rho} \gamma^{\theta} D_{\rho} \bar{F^\nu} _\theta+\frac{1}{\sqrt{2}} \gamma^{\theta} \gamma^{\phi} D^{\nu} \bar{F}_{\theta \phi} \Big\} \psi_{\nu}.
	\end{split}
\end{align}
\subsection*{Trace of $E$}
\begin{equation}\label{trace E}
	\text {tr}\thickspace(E)=\text{tr}\thickspace({E^{\psi_{\mu}}}_{\psi_{\mu}}+{E^\lambda} _\lambda+ {E^{\psi_{\mu}}}_{\lambda}+{E^{\lambda}}_{\psi_{\mu}}).
\end{equation}
From \eqref{E Appendix}, we have
\begin{equation}\label{componentwise trace E}
	\begin{split}
		\text{tr} \thickspace ({E^{\psi_{\mu}}}_{\psi_{\mu}})&=-4\bar{F}^{\mu\nu}\bar{F}_{\mu\nu},
		\\
		\text{tr} \thickspace ({E^{\lambda}}_{\lambda})&=-4\bar{F}^{\mu\nu}\bar{F}_{\mu\nu},
	\end{split}
	\hspace{0.3in}
	\begin{split}
		\text{tr}\thickspace ({E^{\psi_{\mu}}}_{\lambda})&=0,
		\\
		\text{tr} \thickspace ({E^{\lambda}}_{\psi_{\mu}})&=0.
	\end{split}
\end{equation}
\Cref{componentwise trace E,trace E} gives the $\text{tr} \thickspace (E)$ \eqref{Trace}.
\subsection*{Trace of $E^2$}
\begin{align}\label{trace of E2}
	\text{tr}\thickspace (E^2)=\text{tr}\thickspace\left({E^{\psi_{\mu}}}_{\psi_{\nu}} {E^{\psi_{\nu}}}_{\psi_{\mu}}+{E^{\lambda}}_{\lambda} {E^{\lambda} }_{\lambda}+{E^{\psi_{\mu}}}_{\lambda} {E^\lambda}_{\psi_{\mu}}+{E^\lambda}_{\psi_{\mu}} {E^{\psi_{\mu}}}_{\lambda}\right).
\end{align}
Using \cref{E Appendix}, we have
\begin{align}\label{component wise E2 trace}
	\begin{split}
		\text{tr} \thickspace ({E^{\psi_{\mu}}}_{\psi_{\nu}}{E^{\psi_{\nu}}}_{\psi_{\mu}}) &=5R^{\mu\nu}R_{\mu\nu}+2(\bar{F}^{\mu\nu}\bar{F}_{\mu\nu})^2 +2R^{\mu\nu\theta\phi}R_{\mu\nu\theta\phi},\\
		\text{tr} \thickspace ({E^{\lambda}}_{\lambda}{E^{\lambda}}_{\lambda}) &=8(\bar{F}^{\mu\nu} \bar{F}_{\mu\nu})^2-4R^{\mu \nu}R_{\mu\nu},\\
		\text{tr} \thickspace ({E^{\psi_{\mu}}}_{\lambda} {E^{\lambda}}_{\psi_{\mu}})&=R^{\mu\nu}R_{\mu\nu}-R_{\mu\nu\theta\phi}\bar{F}^{\mu\nu}\bar{F}^{\theta \phi},\\
		\text{tr} \thickspace ({E^{\lambda}}_{\psi_{\mu}}{E^{\psi_{\mu}}}_{\lambda})&=R^{\mu\nu}R_{\mu\nu}-R_{\mu\nu\theta\phi}\bar{F}^{\mu\nu}\bar{F}^{\theta \phi}.
	\end{split}
\end{align}
\Cref{component wise E2 trace,trace of E2} yields the $\text{tr} \thickspace (E^2)$ \eqref{Trace}.
\subsection*{Trace of $\Omega_{\alpha\beta}\Omega^{\alpha\beta}$}
The expression for $\Omega_{\alpha \beta}$ using $\omega_{\rho}$ \eqref{omega 2} is
\begin{align}\label{Omega Appendix}
	\begin{split}
		\tilde{\varphi}_{m}(\Omega_{\alpha\beta})^{mn}\tilde{\varphi}_{n}
		=&\bar{\psi}_{\mu}(\Omega_{\alpha \beta})^{\psi_{\mu}\psi_{\nu}}\psi_{\nu}+\bar{\lambda}(\Omega_{\alpha \beta})^{\lambda \lambda}\lambda+\bar{\psi}_{\mu}(\Omega_{\alpha \beta})^{\psi_{\mu} \lambda}\lambda+\bar{\lambda}(\Omega_{\alpha \beta})^{\lambda \psi_{\nu}}\psi_{\nu}
		\\
		=&\bar{\psi}_{\mu}\Big\{\mathbb{I}_{4}{R^{\mu\nu}}_{\alpha\beta}+\frac{1}{4}\bar{g}^{\mu\nu}\gamma^{\xi} \gamma^{\kappa}R_{\alpha \beta \xi \kappa}+\left[\omega_{\alpha},\omega_{\beta}\right]^{\psi_{\mu}\psi_{\nu}}  \Big\}\psi_{\nu}
		\\
		&+\bar{\lambda}\Big\{\frac{1}{4} \gamma^{\theta}\gamma^{\phi} R_{\alpha \beta \theta \phi}+[\omega_{\alpha},\omega_{\beta}]^{\lambda\lambda}\Big\}\lambda
		\\
		&+\bar{\psi}_{\mu}\Big\{{D_{[\alpha}\omega_{\beta]}}^{\psi_{\mu}\lambda} \Big\}\lambda+\bar{\lambda}\Big\{{D_{[\alpha}\omega_{\beta]}}^{\lambda\psi_{\nu}} \Big\}\psi_{\nu}.
	\end{split}
\end{align}
One can calculate $\text{tr}\thickspace (\Omega_{\alpha\beta}\Omega^{\alpha\beta})$ as
\begin{align}
	\begin{split}\label{Trace Omega2}
		\text{tr} \thickspace (\Omega^{\alpha \beta}\Omega_{\alpha \beta})=&\text{tr}\thickspace \Big\{{(\Omega_{\alpha \beta})_{\psi_{\mu}}}^{\psi_{\nu}} {(\Omega^{\alpha \beta})_{\psi_{\nu}}}^{\psi_{\mu}}+{(\Omega_{\alpha \beta})_{\lambda}}^{\lambda}{(\Omega^{\alpha \beta})_{\lambda}}^{\lambda} 
		\\&
		+{(\Omega_{\alpha \beta})_{\psi_{\mu}}}^{\lambda}{(\Omega^{\alpha \beta})_{\lambda}}^{\psi_{\mu}}+{(\Omega_{\alpha\beta})_{\lambda}}^{\psi_{\nu}}{(\Omega^{\alpha \beta})_{\psi_{\nu}}}^{\lambda}\Big\},
	\end{split}
\end{align}
where
\allowdisplaybreaks{
	\begin{align}\label{Term 1}
		\begin{split}
			&{(\Omega_{\alpha \beta})_{\psi_{\mu}}}^{\psi_{\nu}} {(\Omega^{\alpha \beta})_{\psi_{\nu}}}^{\psi_{\mu}}
			\\
			=&\Big\{\underbrace{{{\mathbb{I}_{4}R_{\mu}}^{\nu}}_{\alpha \beta}}_{R_1}+\underbrace{\frac{1}{4}\bar{g_{\mu}}^{\nu}\gamma^{\xi} \gamma^{\kappa}R_{\alpha \beta \xi \kappa}}_{R_2}+\Big(\underbrace{{(\omega_{\alpha})_{\psi_{\mu}}}^{\lambda}{(\omega_{\beta})_{\lambda}}^{\psi_{\nu}}-{(\omega_{\beta})_{\psi_{\mu}}}^{\lambda}{(\omega_{\alpha})_{\lambda}}^{\psi_{\nu}}}_{R_3}\Big) \Big\}
			\\
			&\times\Big\{\underbrace{\mathbb{I}_{4}{{R_{\nu}}^{\mu}}^{\alpha \beta}}_{S_1}+\underbrace{\frac{1}{4}\bar{g_{\nu}}^{\mu}\gamma^{\sigma} \gamma^{\kappa}{R^{\alpha \beta}}_{\sigma \kappa}}_{S_2}+\Big(\underbrace{{(\omega^{\alpha})_{\psi_{\nu}}}^{\lambda}{(\omega^{\beta})_{\lambda}}^{\psi_{\mu}}-{(\omega^{\beta})_{\psi_{\nu}}}^{\lambda}{(\omega^{\alpha})_{\lambda}}^{\psi_{\mu}}}_{S_3}\Big)\Big\},
		\end{split}
	\end{align}
	\begin{align}\label{Term 2}
		\begin{split}
			{(\Omega_{\alpha \beta})_{\lambda}}^{\lambda}{(\Omega^{\alpha \beta})_{\lambda}}^{\lambda}=&\Big\{\underbrace{\frac{1}{4}\gamma^{\theta}\gamma^{\phi}R_{\alpha\beta\theta\phi}}_{R_4}+\underbrace{{(\omega_{\alpha})_{\lambda}}^{\psi_{\mu}}{(\omega_{\beta})_{\psi_{\mu}}}^{\lambda}-{(\omega_{\beta})_{\lambda}}^{\psi_{\mu}}{(\omega_{\alpha})_{\psi_{\mu}}}^{\lambda}}_{R_5} \Big\}
			\\
			&\times\Big\{\underbrace{\frac{1}{4}\gamma^{\rho}\gamma^{\sigma}{R^{\alpha\beta}}_{\rho\sigma}}_{S_4}+\underbrace{{(\omega^{\alpha})_{\lambda}}^{\psi_{\nu}}{(\omega^{\beta})_{\psi_{\nu}}}^{\lambda}-{(\omega^{\beta})_{\lambda}}^{\psi_{\nu}}{(\omega^{\alpha})_{\psi_{\nu}}}^{\lambda}}_{S_5}\Big\},
		\end{split}
	\end{align}
	\begin{align}\label{Term 3}
		{(\Omega_{\alpha \beta})_{\psi_{\mu}}}^{\lambda}{(\Omega^{\alpha \beta})_{\lambda}}^{\psi_{\mu}}=\underbrace{\Big(D_{\alpha}{(\omega_{\beta})_{\psi_{\mu}}}^{\lambda}-D_{\beta}{(\omega_{\alpha})_{\psi_{\mu}}}^{\lambda}\Big)}_{R_6}\times\underbrace{\Big(D^{\alpha}{(\omega^{\beta})_{\lambda}}^{\psi_{\mu}}-D^{\beta}{(\omega^{\alpha})_{\lambda}}^{\psi_{\mu}}\Big)}_{S_6},
\end{align}}
and
\begin{align}\label{Term 4}
	{(\Omega_{\alpha\beta})_{\lambda}}^{\psi_{\nu}}{(\Omega^{\alpha \beta})_{\psi_{\nu}}}^{\lambda}=\underbrace{\Big(D_{\alpha}{(\omega_{\beta})_{\lambda}}^{\psi_{\nu}}-D_{\beta}{(\omega_{\alpha})_{\lambda}}^{\psi_{\nu}}\Big)}_{R_7}\times\underbrace{\Big(D^{\alpha}{(\omega^{\beta})_{\psi_{\nu}}}^{\lambda}-D^{\beta}{(\omega^{\alpha})_{\psi_{\nu}}}^{\lambda}\Big)}_{S_7}.
\end{align}
Then,
\begin{equation}\label{componentwise trace result}
	\begin{split}
		\text{tr}\thickspace (R_1 S_1) &=-4R^{\mu\nu\theta\phi}R_{\mu\nu\theta\phi},\\
		\text{tr}\thickspace (R_2 S_2) &=-2R^{\mu\nu\theta\phi}R_{\mu\nu\theta\phi},\\
		\text{tr}\thickspace (R_1 S_2) &=0,\\
		\text{tr}\thickspace (R_2 S_1) &=0,\\
		\text{tr}\thickspace (R_1 S_3) &=-4R_{\mu\nu\theta\phi}\bar{F}^{\mu\nu}\bar{F}^{\theta \phi},\\
		\text{tr}\thickspace (R_3 S_1) &=-4R_{\mu\nu\theta\phi}\bar{F}^{\mu\nu}\bar{F}^{\theta \phi},\\
		\text{tr}\thickspace (R_2 S_3)&=2R^{\mu\nu}R_{\mu\nu},\\
		\text{tr}\thickspace (R_3 S_2)&=2R^{\mu\nu}R_{\mu\nu},\\
	\end{split}
	\hspace{0.3in}
	\begin{split}
		\text{tr}\thickspace (R_3 S_3)&=6R^{\mu\nu}R_{\mu\nu}-36 (\bar{F}^{\mu\nu}\bar{F}_{\mu\nu})^2,\\
		\text{tr} \thickspace (R_{4}S_{4})&=-\frac{1}{2}R^{\mu\nu\theta\phi}R_{\mu\nu\theta\phi},\\
		\text{tr} \thickspace (R_4 S_{5})&=-2R^{\mu\nu}R_{\mu\nu},\\
		\text{tr} \thickspace (R_5 S_{4}) &=-2R^{\mu\nu}R_{\mu\nu},\\
		\text{tr}\thickspace (R_5 S_5) &=8R^{\mu\nu}R_{\mu\nu}-24(\bar{F}^{\mu\nu}\bar{F}_{\mu\nu})^2,\\
		\text{tr}\thickspace (R_6 S_6) &=10 R_{\mu\nu\theta\phi}\bar{F}^{\mu\nu}\bar{F}^{\theta \phi}-10 R^{\mu\nu}R_{\mu\nu},\\
		\text{tr}\thickspace (R_7 S_7) &=10 R_{\mu\nu\theta\phi}\bar{F}^{\mu\nu}\bar{F}^{\theta \phi}-10 R^{\mu\nu}R_{\mu\nu}.
	\end{split}
\end{equation}
\Cref{componentwise trace result,Term 2,Term 1,Term 3,Term 4} gives
\begin{align}\label{result}
	\begin{split}
		\text{tr}\thickspace ({(\Omega_{\alpha \beta})_{\psi_{\mu}}}^{\psi_{\nu}} {(\Omega^{\alpha \beta})_{\psi_{\nu}}}^{\psi_{\mu}}) &=-6R^{\mu\nu\theta\phi}R_{\mu\nu\theta\phi}-8R_{\mu\nu\theta\phi}\bar{F}^{\mu\nu}\bar{F}^{\theta \phi}\\
		&\qquad +10 R^{\mu\nu}R_{\mu\nu}-36 (\bar{F}^{\mu\nu}\bar{F}_{\mu\nu})^2,\\
		\text{tr}\thickspace({(\Omega_{\alpha \beta})_{\lambda}}^{\lambda}{(\Omega^{\alpha \beta})_{\lambda}}^{\lambda}) &=-\frac{1}{2}R^{\mu\nu\theta\phi}R_{\mu\nu\theta\phi}+4R^{\mu\nu}R_{\mu\nu}-24 (\bar{F}^{\mu\nu}\bar{F}_{\mu\nu})^2,\\
		\text{tr}\thickspace({(\Omega_{\alpha \beta})_{\psi_{\mu}}}^{\lambda}{(\Omega^{\alpha \beta})_{\lambda}}^{\psi_{\mu}})
		&=10 R_{\mu\nu\theta\phi}\bar{F}^{\mu\nu}\bar{F}^{\theta \phi}-10 R^{\mu\nu}R_{\mu\nu},\\
		\text{tr}\thickspace ({(\Omega_{\alpha\beta})_{\lambda}}^{\psi_{\nu}}{(\Omega^{\alpha \beta})_{\psi_{\nu}}}^{\lambda}) &=10 R_{\mu\nu\theta\phi}\bar{F}^{\mu\nu}\bar{F}^{\theta \phi}-10 R^{\mu\nu}R_{\mu\nu}. 
	\end{split}
\end{align}
Finally, from \eqref{Trace Omega2} and \eqref{result}, we have obtained $\text{tr}\thickspace (\Omega_{\alpha \beta}\Omega^{\alpha \beta})$ \eqref{Trace}.
%\end{ceqn}

\end{document}